\documentclass[%
preprint,
superscriptaddress,
unsortedaddress,
 amsmath,amssymb,
 aps,
]{revtex4-1}

\usepackage{graphicx}
\usepackage{dcolumn}
\usepackage{bm}
\usepackage[mathlines]{lineno}
\usepackage{xcolor}
\usepackage{xr}
\usepackage[normalem]{ulem}
\usepackage{natbib}
\usepackage{bm}
\usepackage[mathlines]{lineno}
\usepackage{float}
\usepackage{tikz}
\usepackage{xcolor}

\newcommand{\jgb}[1]{{\color{black}{#1}}}  
\newcommand{\pn}[1]{{ \color{black}{#1}}}  
\newcommand{\yas}[1]{{ \color{black}{#1}}}  
\newcommand{\ip}{{i^\prime}}
\newcommand{\jp}{{j^\prime}}

\begin{document}

\title{Interactions between Large Molecules: Puzzle for Reference Quantum-Mechanical Methods
}

\author{Yasmine S. Al-Hamdani$^{\ddagger}$}
\affiliation{Department of Chemistry, University of Zurich, CH-8057 Z\"urich, Switzerland}
\affiliation{Department of Physics and Materials Science, University of Luxembourg, L-1511 Luxembourg, Luxembourg}
\author{P\'eter R. Nagy$^{\ddagger}$}%
\affiliation{
Department of Physical Chemistry and Materials Science, Budapest University of Technology and Economics, H-1521 Budapest, P.O.Box 91, Hungary}
\author{Dennis Barton}
\affiliation{Department of Physics and Materials Science, University of Luxembourg, L-1511 Luxembourg, Luxembourg}
\author{Mih\'aly K\'allay}%
\affiliation{
Department of Physical Chemistry and Materials Science, Budapest University of Technology and Economics, H-1521 Budapest, P.O.Box 91, Hungary}
\author{Jan Gerit Brandenburg}
\email{j.g.brandenburg@gmx.de}
\affiliation{
Interdisciplinary Center for Scientific Computing,
University of Heidelberg, Im Neuenheimer Feld 205A, 69120 Heidelberg, Germany}
\affiliation{
Digital Organization,
Merck KGaA, Frankfurter Str. 250, 64293 Darmstadt, Germany}
\author{Alexandre Tkatchenko}
\email{alexandre.tkatchenko@uni.lu}
\affiliation{Department of Physics and Materials Science, University of Luxembourg, L-1511 Luxembourg, Luxembourg }
\author{
$^{\ddagger}$These authors contributed equally.
}



\begin{abstract}
{\bf
Quantum-mechanical methods are widely used for understanding molecular interactions throughout biology, chemistry, and materials science. 
Quantum diffusion Monte Carlo (DMC) and coupled cluster with single, double, and perturbative triple excitations [CCSD(T)] are two state-of-the-art and trusted wavefunction methods that have been categorically shown to yield accurate interaction 
energies for small\pn{organic} molecules.
\yas{These methods provide valuable reference information for widely-used semi-empirical and machine learning potentials, especially where experimental information is scarce.}
However, agreement for systems beyond small molecules is a crucial remaining milestone for cementing the benchmark accuracy of these methods. Approaching such well-converged predictive power in larger molecules has motivated major developments in CCSD(T) as well as DMC algorithms in the past years, resulting in orders of magnitude time-to-solution reductions. 
Here, we show that  CCSD(T) and DMC interaction energies are not in consistent agreement for a set of polarizable supramolecules. 
Whilst agreement is found for some of the complexes, in a few key systems disagreements of up to 8 kcal mol$^{-1}$ remained.
This leads to differences of up to 6 orders of magnitude in the corresponding binding association constant at room temperature \pn{for systems which are well within 
the accustomed domain of applicability for both methods.}
\pn{These findings thus indicate that more caution is required 
when aiming at reproducible non-covalent interactions between extended molecules.}
Our data contradicts the expectation that the most comprehensive and robust wavefunction methods predict identical non-covalent interactions and indicate \jgb{an unsolved challenge} for benchmark approaches.  
} 
\end{abstract}

\maketitle

\section*{Main}

The most accurate methods for studying matter at the atomic scale are wavefunction-based approaches which explicitly account for many-electron interactions. Given only the positions and nuclear charges of atoms, we can now predict, among basically every observable property, the binding strength of relatively small molecular systems (\textit{i.e.} less than 50 atoms) to within a few tenths of a kcal mol$^{-1}$ using many-body solutions to the Schr\"{o}dinger equation~\cite{Carter2008,Dubecky2013a,Chan-Science}. 
This value is better than the so-called ``chemical accuracy" of 1 kcal mol$^{-1}$ required for reliable predictions of thermodynamic properties. Indeed, the relative stabilities of many non-covalently bound materials such as 2D layered materials, pharmaceutical drugs, and different polymorphs of ice, are underpinned by small energy differences on the order of tenths of a kcal mol$^{-1}$~\cite{Reilly2016}. 
However, experimentally determining binding affinities under well-defined, pristine conditions is notoriously challenging~\cite{Muller-Dethlefs2000}.
In addition, thousands of computational works describe physical interactions in materials, which are not well understood at the experimental level, for instance, as part of 
rational design initiatives in novel materials 
including soft colloidal matter, nanostructures, metal organic and covalent organic frameworks~\cite{Wang2018,Lee2018,Ongari2019}.
The present shortage of benchmark information is a major setback for forming reliable predictions across the natural sciences and is frequently addressed through demanding, but increasingly feasible, wavefunction-based methods. However, extending the use of highly-accurate methods to a regime of larger molecules is hindered by theoretical and technical challenges due to the steep increase in computational cost required for an accurate description of many-electron interactions~\cite{Al-Hamdani2019,CCSD_DMC_solidH}.
\begin{figure}[b]
\includegraphics[width=0.45\textwidth]{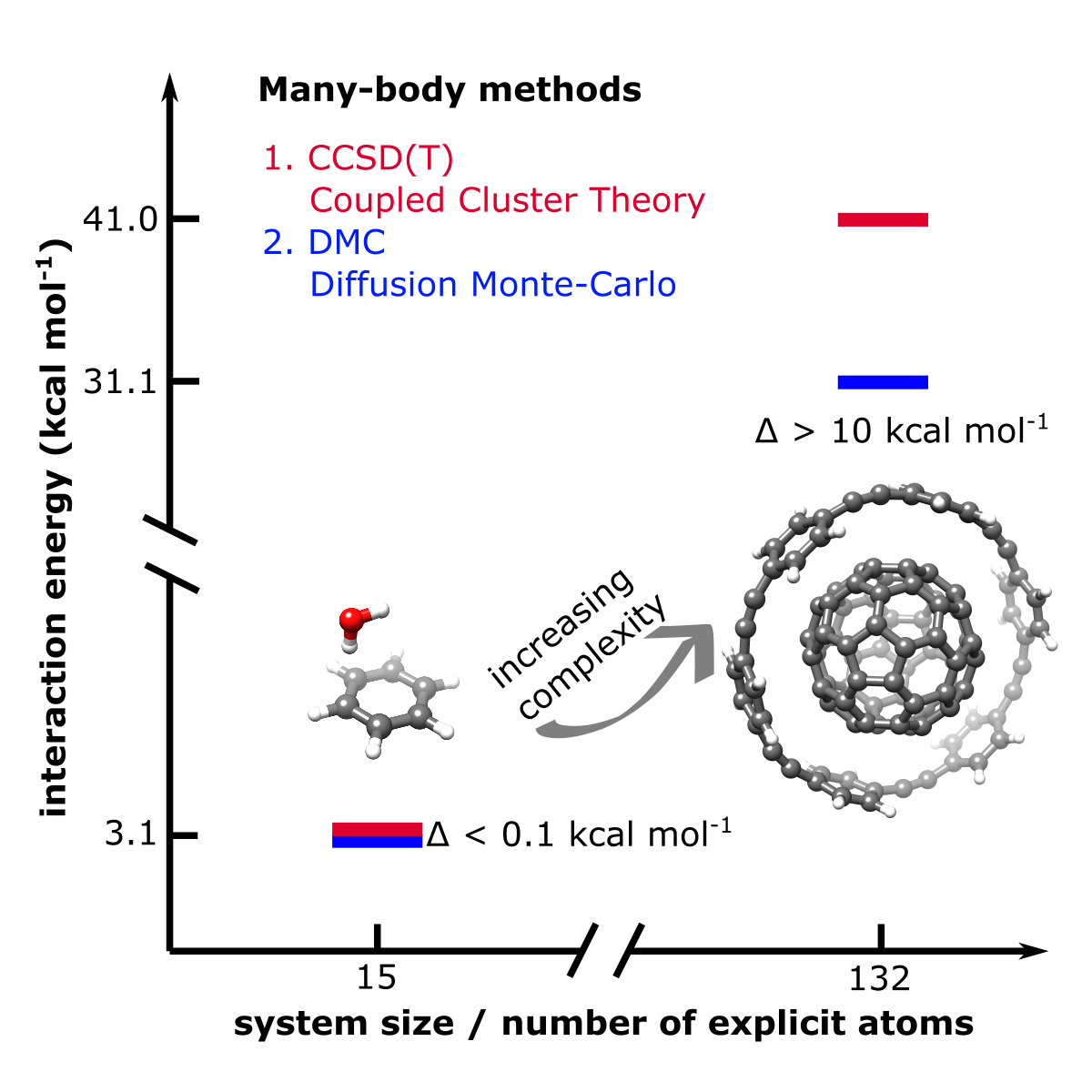}
\caption{\label{fig1:toc} The CCSD(T) and FN-DMC computed binding energies of a water-benzene dimer~\cite{waterongraphene} is shown in comparison to a buckyball-ring complex computed here. It can be seen that the binding energy increases by a factor $\sim10$, near-linearly with the size of the system, whereas the corresponding disagreement between CCSD(T) and FN-DMC increases by a factor of $\sim100$.}
\end{figure}

Here we use two widely trusted wavefunction methods that can provide sub-chemically accurate solutions to the electronic Schr{\"o}dinger equation for non-covalent interactions.
First, we utilize coupled-cluster (CC) theory with single, double, and perturbative triple excitations [CCSD(T)]~\cite{CCSD(T)} -- approximated via the local 
natural orbital (LNO) scheme to be practicable [LNO-CCSD(T)]~\cite{LocalCC3,LocalCC4}. Coupled cluster theory has gained great prominence in the last 30 years and the label of `gold-standard' for remarkable accuracy on virtually all systems
in its domain of applicability~\cite{ShavittBartlettBook}.
Second, a stochastic quantum method that computes the energy for the many-electron wavefunction directly is known as fixed-node diffusion Monte Carlo (FN-DMC). This method has seen a surge of use in recent years, particularly for predicting large molecules and periodic systems with non-covalent interactions~\cite{Dubecky2016,CCSD_DMC_solidH}, such as molecular crystals and adsorption on 2D materials~\cite{Dubecky2016,Zen2018,al-hamdani2017cnt}.

As we demonstrate in Fig.~\ref{fig1:toc}, CCSD(T) and FN-DMC \pn{interaction energies} are in sub-chemical agreement in small systems such as the benzene-water dimer~\cite{waterongraphene}.
Nonetheless, FN-DMC and CCSD(T) are still prohibitively expensive for most applications in biology and chemistry, and as result, 
very little is known about how predictive these theoretical methods are in the regime of larger molecules.

\begin{figure}[hbt]
\includegraphics[width=8cm]{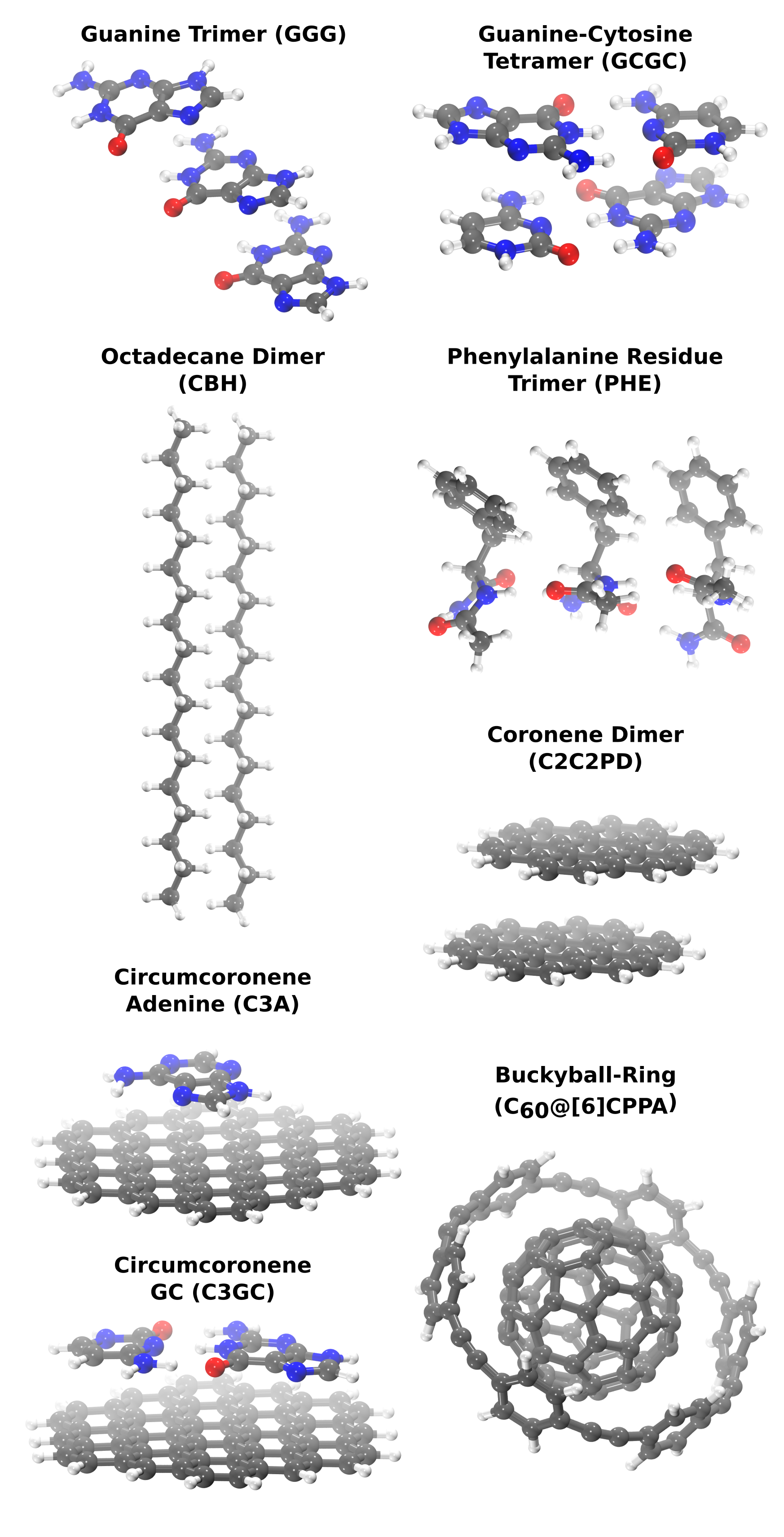}
\caption{\label{fig1:intro} The supramolecular complexes from L7 data set~\cite{L7set} and a buckyball-ring supramolecular complex consisting of 132 atoms.}
\end{figure}

Straightforward extrapolations of interactions from small molecules to large complexes are difficult to make due to the interplay and accumulation of interactions that are non-additive, anisotropic, or have many-body character~\cite{Ambrosetti-Science,Jordan2019,Jenness2010a,waterongraphene,mp2_rpa_nci_fruche}. 
As such, a deeper understanding of non-covalent interactions can be gained by directly applying state-of-the-art methods in larger molecular complexes. Here, we use a frequently studied compilation, the L7 molecular data set from Sedlak \textit{et al.}~\cite{L7set} to ascertain the predictive power of FN-DMC and CCSD(T) for relatively large complexes involving intricate $\pi-\pi$ stacking, electrostatic interactions, and hydrogen-bonding (see Fig.~\ref{fig1:intro}). 
In addition, we consider a larger system of a C$_{60}$ buckyball inside a [6]-cycloparaphenyleneacetylene ring (which we label as C$_{60}$@[6]CPPA), consisting of 132 atoms. 
This structure has a number of interesting features: (i) an open-framework that can be found in covalent organic frameworks and carbon nanotubes, (ii) the buckyball has a large polarizability ($76\pm8$~\AA$^3$)~\cite{Antoine1999}
which gives \yas{rise} to considerable dispersion interactions, and (iii) confinement between the ring and the buckyball that may cause non-trivial long-range repulsive interactions~\cite{Sadhukhan2017}. 

Following recent algorithmic advances for more efficient CCSD(T) and FN-DMC, we predict interaction energies for a set of supramolecular complexes and converge numerical thresholds to the best of our joint knowledge and expertise. 
\yas{Hereafter, we refer to CCSD(T) and FN-DMC interaction energies but note that a number of approximations are used in both methods. More specifically, the CCSD(T) interaction energies we report come from systematically converging LNO-CCSD(T) towards canonical CCSD(T). Meanwhile, the significance of approximations in FN-DMC interaction energies are assessed using statistical measures.}
CCSD(T) and FN-DMC \pn{are consistent} for five of the eight supramolecular complexes, covering a range of interactions including hydrogen bonding and $\pi-\pi$ stacking. 
However, we find that three key 
complexes reveal several kcal mol$^{-1}$ differences between best estimated CCSD(T) and FN-DMC calculations. 
Most notably, a substantial disagreement of $\sim$8 kcal mol$^{-1}$ (or $20~\%$) is found in the interaction energy ($E_\text{int}$ as defined in Methods)
of the buckyball-ring system. \pn{This 8 kcal mol$^{-1}$ inconsistency remains on top of the uncertainty estimates incorporating all controllable 
sources of errors.
We also gauge the impact of approximations intrinsic to each method, not covered in the numerical uncertainty estimates, and find that 8 kcal mol$^{-1}$ is an order of magnitude beyond these.
It is thus yet unclear whether this discrepancy would also be present between the approximation-free CCSD(T) and DMC results or it is a result of an unexplored source of error.}
 \pn{As shown in Fig. \ref{fig1:toc} and in Table \ref{tab:my-table} below, such a sizable deviation cannot 
be explained solely by the size-extensive growth of the difference between CCSD(T) and FN-DMC.}
Consequently, the interaction energies of three of the complexes considered here are still unsettled.

We applied two different, widely-used and well-performing DFT 
approaches developed for capturing long-range \yas{dispersion interactions: DFT+D4~\cite{dftd4} and DFT+MBD~\cite{MBD}. Both methods model London dispersion based on a coarse-grained description and account for all orders of many-body dispersion in  different manner. See Refs.~\cite{Hermann2017a,Grimme2016} for an overview of various ways to capture dispersion in the DFT framework.}
We find that DFT+MBD closely matches FN-DMC, while the recent DFT+D4 method agrees well with CCSD(T), irrespective of the level of disagreement between CCSD(T) and FN-DMC.
Therefore, the absence of either CCSD(T) or FN-DMC references could incorrectly suggest that one of the DFT methods performs better than the other.
This illustrates that the unprecedented level of disagreement amongst state-of-the-art methods in large organic molecules has consequences well outside the developer communities.

\subsection*{State-of-the-art methods for non-covalent interactions}
CCSD(T) and FN-DMC methods account for dynamic electron correlation 
through an expansion in electron configurations
in the former and through the real-space fluctuation of electrons in the latter.
These two equally viable formulations can be \pn{illustrated by} the corresponding expressions of $\Psi({\bf R)}$, the exact wavefunction:
\begin{itemize}
    \item [1.] {\bf DMC:} Imaginary time ($\tau$) propagation of a trial function $\Psi_{\mathrm{T}}({\bf R})$ in real space: 

    $\displaystyle|\Psi({\bf R}) \rangle  = \lim_{\tau \to \infty}  \text{exp}\!\left[-\tau(\hat{H}-E_\mathrm{T})\right] |\Psi_{\mathrm{T}}({\bf R}) \rangle$ 
    \item[2.] {\bf CC:} Expansion of excited determinants generated via the operator $\hat{T}_n$ from a reference wavefunction: $\displaystyle|\Psi({\bf R}) \rangle  =   \text{exp}\!\left[\sum_{n=1}^{\infty}\hat{T}_n\right] |\Psi_{\mathrm{T}}({\bf R}) \rangle$
\end{itemize}
The crucial challenge lies in extensively accounting for relatively small fluctuations in the electron charge densities. In FN-DMC this implies the need for relatively small time-steps $\Delta\tau$ for the projection of the wavefunction as well as an extensive sampling of electron configurations in real-space ($\lim_{\tau \to \infty}$) in order to reduce the stochastic noise in the predicted energy. 
In coupled cluster theory, non-covalent interactions require a high-order treatment of many-electron processes, as is included in CCSD(T), and a sufficiently large single-particle basis set. Reaching basis set saturation and well-controlled local approximations concurrently for the studied systems required previously unfeasible computational efforts as shown by the several kcal mol$^{-1}$ scatter of interaction energy predictions reported for the L7 set (see Fig. \ref{fig2:mainres}).
Our recent efforts enabled the following: 
(i) a systematically
converging series of local CCSD(T) results is presented for highly-complicated complexes, 
(ii) both the local and the
basis set incompleteness (BSI) errors are closely monitored using comprehensive uncertainty measures~\cite{LocalCC4},
(iii) convergence up to chemical accuracy is reached for the complete L7 set  
concurrently in the local approximations as well as in the basis set saturation.
The benefit of such demanding convergence studies is that the resulting interaction energies, up to the respective error bars, 
can be considered independent of the \pn{corresponding}
approximations. 
Consequently, we expect that 
the CBS limit of the exact CCSD(T) results could, in principle, be approached similarly  using alternative basis sets~\cite{QMC_CC_solids,CCSD_DMC_solidH,MRA_CC2} or local correlation methods~\cite{DLPNO-CCSD(T),PNOCCreview,PNO-CCSDHattig,DLPNO-CCSD(T)-F12}, as it is clearly 
observed for some of the present complexes (see, \textit{e.g.} GGG or CBH in Fig. \ref{fig2:mainres}).

We use highly-optimized algorithms both for FN-DMC and CCSD(T) as outlined in Methods, and push them beyond the typically applied limits. 
We used \textit{circa} 0.7 and 1 million CPU core hours for 
FN-DMC and CCSD(T), respectively. This is equivalent to running a modern 28 core machine constantly for $\sim7$ years.

\subsection*{Losing consensus on supramolecular interactions} 
Demonstrating agreement between fundamentally different electronic structure methods for solving the Schr\"{o}dinger equation provides a proof-of-principle for the accuracy of the methods beyond technical challenges. 
\pn{To date, disagreements between CCSD(T) and FN-DMC have been reported only for systems where their key assumptions, \textit{e.g.} single-reference wavefunction, accurate node-structure, etc. were not completely 
fulfilled~\cite{Be2_dmc,dlpno_spincrossover}. 
Previously however, CCSD(T) and FN-DMC were found in agreement within the error bars, 
for the interaction energies of small organic molecules with pure dynamic correlation~\cite{A24set2,Dubecky2016,Al-Hamdani2019} as well as some extended systems~\cite{waterongraphene,Zen2018,lih_dmc_cc}.
Establishing this agreement for systems at the 100 atom range has, however, been hindered by the sizable or unavailable error estimates for finite systems~\cite{Al-Hamdani2019}. }
\yas{For example, binding energies of large host-guest complexes derived from experimental association free energies~\cite{s12l,Sure2015} motivated previous
FN-DMC~\cite{Hermann2017} as well as local CCSD(T)~\cite{Calbo2015} computations. However, conclusive remarks could not be made on the consistency of FN-DMC and local CCSD(T) on these
complexes due to technical difficulties and unavailable uncertainty estimates for local CCSD(T), and large error estimates on both experimental and FN-DMC energies (up to a few kcal mol$^{-1}$).
}

\begin{figure}[htb]
\includegraphics[width=8.6cm]{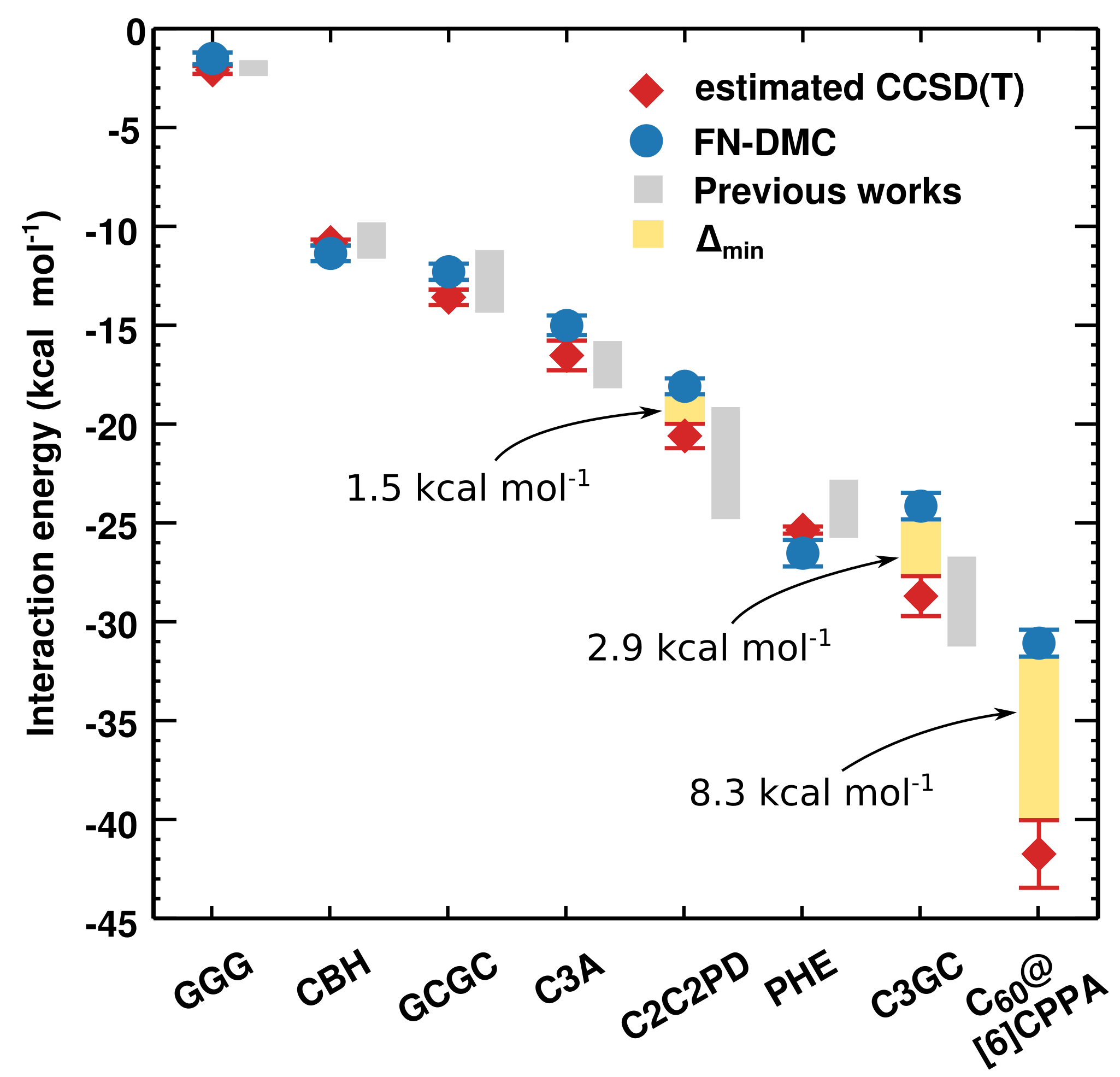}
\caption{\label{fig2:mainres} CCSD(T) and FN-DMC interaction energies for the supramolecular complexes 
of the L7 data set~\cite{L7set} and the C$_{60}$@[6]CPPA buckyball-ring complex arranged in terms of increasing interaction strength.
Gray bars mark the range of interaction energies reported in the literature using alternative wavefunction based methods [\textit{e.g.} QCISD(T)~\cite{L7set},
and various local CCSD(T) approaches~\cite{DHNLsupramol,L7CCGrimme1,B97-3c,XSAPT-MBD,DLPNO-CCSD(T)-F12,PNOCCreview}].
The yellow bars indicate the delta value ($\Delta_\text{min}$) which is the minimum difference between best converged CCSD(T) and FN-DMC, given by the estimated and stochastic error bars, respectively.}
\end{figure}

\pn{Here, we consider similar but somewhat smaller supramolecular complexes (Fig.~\ref{fig1:intro}) and
obtain tightly converged local CCSD(T) and FN-DMC results sufficient for rigorous comparisons (see Fig.~\ref{fig2:mainres} and Table~\ref{tab:my-table}).}
The level of uncertainty in our results is indicated by stochastic error bars for FN-DMC and the sum of local and BSI error estimates for CCSD(T). 
The complexes are arranged in Fig. \ref{fig2:mainres} according to increasing interaction strength, which roughly scales with the size of the interacting surface. 
CCSD(T) and FN-DMC agree on the interaction energy \yas{to within 0.5 kcal mol$^{-1}$, taking error bars into account,} for a subset of the complexes we consider: GGG, CBH, GCGC, C3A and PHE. These complexes are between 48 and 112 atoms in size and exhibit $\pi-\pi$ stacking, hydrogen bonding, and dispersion interactions. Therefore, the agreement for these five complexes indicates their \yas{absolute} interaction energies are established references and can be used to benchmark other methods for large molecules.
\yas{Here, relative differences of very small interaction energies have to be interpreted carefully as they are sensitive to the uncertainty estimates. In GGG for example, $\Delta_\text{min}$ is 0.1 kcal mol$^{-1}$ whilst the relative disagreement lies between 3$\%$ and 50$\%$.
In contrast, the relative disagreement between FN-DMC and CCSD(T) is better resolved in the more strongly interacting C$_{60}$@[6]CPPA complex, at 20--31$\%$.}

\begin{table}[htb!]
\caption{Interaction energies  in kcal mol$^{-1}$ for best estimated CCSD(T) and FN-DMC\pn{, as well as their minimum differences ($\Delta_\text{min}$)} 
for the L7 supramolecular data set and the buckyball-ring complex (C$_{60}$@[6]CPPA).
The indicated errors for CCSD(T) are extrapolated from the convergence of basis sets and local approximations in LNO-CCSD(T). The errors indicated in FN-DMC interaction energies are stochastic errors, with 1 standard deviation (1-$\sigma$). }
\label{tab:my-table}
\begin{ruledtabular}
\begin{tabular}{c rrrr}
Complex & No. of atoms &  CCSD(T) &  FN-DMC & $\Delta_\text{min}$  \\ \hline
GGG     & 48  & $-2.1 $ $\pm~0.2$ & \yas{$-1.5 $ $\pm~0.3$}  & 0.1 \\
CBH     & 112 & $-11.0$ $\pm~0.2$ & $-11.4$ $\pm~0.4$  & 0.0 \\
GCGC    & 58  & $-13.6$ $\pm~0.4$ & \yas{$-12.3$ $\pm~0.3$}  & 0.5 \\
C3A     & 87  & $-16.5$ $\pm~0.8$ & \yas{$-15.0$ $\pm~0.5$}  & 0.2 \\
C2C2PD  & 72  & $-20.6$ $\pm~0.6$ & $-18.1$ $\pm~0.4$  & 1.5 \\
PHE     & 87  & $-25.4$ $\pm~0.2$ & $-26.5$ $\pm~0.7$  & 0.3 \\
C3GC    & 101 & $-28.7$ $\pm~1.0$ & $-24.2$ $\pm~0.7$  & 2.9 \\
C$_{60}$@[6]CPPA & 132 & $-41.7$ $\pm~1.7$ & $-31.1$ $\pm~0.7$ & 8.3
\end{tabular}%
\end{ruledtabular}
\end{table}

A salient and surprising finding is the disagreement between state-of-the-art methods on the interaction energy of three non-trivial complexes: coronene dimer (C2C2PD), circumcoronene-GC base pair (C3GC), and buckyball-ring (C$_{60}$@[6]CPPA). 
Considering the error bars, the minimum differences ($\Delta_\text{min}$), as indicated in \pn{Table~\ref{tab:my-table} and} 
Fig.~\ref{fig2:mainres} are 1.5, 2.9, and 8.3 kcal mol$^{-1}$ for C2C2PD, C3GC, and C$_{60}$@[6]CPPA, respectively, 
and could be as high as 3.5, 6.2, and 13.1 kcal mol$^{-1}$. 
\pn{Considering the comparable size of C3A, PHE, and CBH to C2C2PD, C3GC, and C$_{60}$@[6]CPPA, the $\Delta_\text{min}$ values of the latter three complexes
are  not explained simply by the large size or  the large area of the interacting surface.}
CCSD(T) predicts consistently stronger interaction in these complexes than FN-DMC,\pn{ but at this point it is unclear what the exact interaction energies are.}

C2C2PD has attracted the most 
attention to date in the CCSD(T) context as it  represents a stepping stone between two widely studied systems: benzene dimer and graphene bilayer~\cite{Al-Hamdani2019}. Already C2C2PD has posed a significant challenge to various local CCSD(T) methods due to its slowly-decaying long-range interactions~\cite{DHNLsupramol,L7CCGrimme1,B97-3c,XSAPT-MBD,DLPNO-CCSD(T)-F12,PNOCCreview,LocalCC4}. Considerable efforts have been devoted recently~\cite{DLPNO-CCSD(T)-F12,PNOCCreview,LocalCC4} to narrow down the local CCSD(T) interaction energy of C2C2PD to the range of about $-19$ to $-21$ kcal mol$^{-1}$. Thus \pn{the presently reported $-20.6$ $\pm~0.6$ kcal mol$^{-1}$ interaction energy
 and  previous local CCSD(T) results, containing 
analogous local approximations,} consistently indicate stronger interaction than FN-DMC for C2C2PD.

\subsection*{Distinct errors using DNA base molecules on circumcoronene}\label{sec:convergence}

The C3GC and C3A complexes are ideal for assessing the convergence of CCSD(T) and FN-DMC, due to their chemical similarity and importance of $\pi-\pi$ stacking interactions, \textit{i.e.} nucleobases stacked on circumcoronene. 
CCSD(T) and FN-DMC \yas{agree within 1 kcal mol$^{-1}$}
for the interaction energy of C3A, whereas there is a notable disagreement of at least 2.9 kcal mol$^{-1}$ in the interaction energy of C3GC. Interestingly, both systems involve similar interaction mechanisms, with C3GC exhibiting both stacking and hydrogen-bonding interactions. 

\begin{figure*}[htb]
\includegraphics[width=16cm]{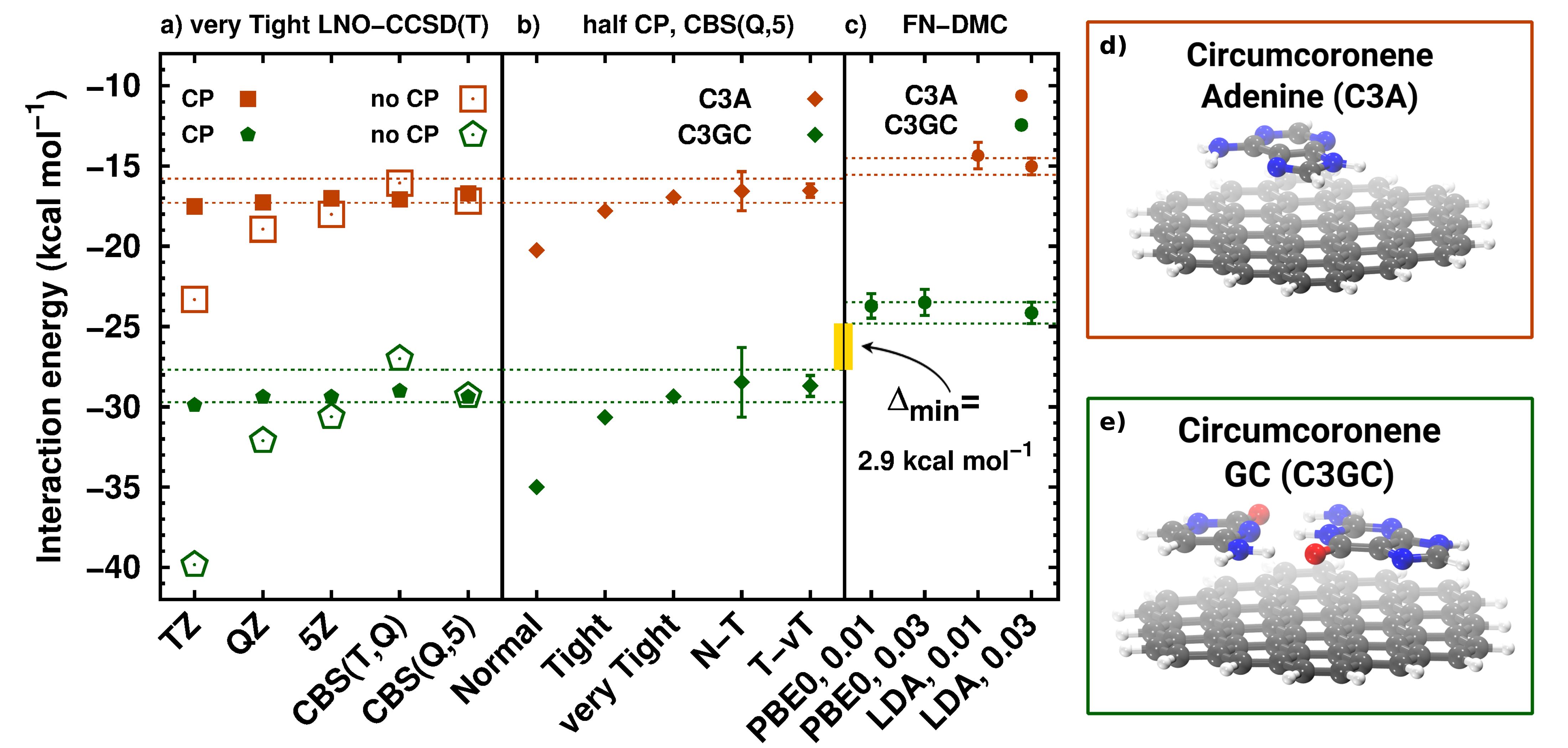}
\caption{\label{fig3:convergence} The interaction energy of the C3A and C3GC complexes using LNO-CCSD(T) (panels a and b) and  FN-DMC (panel c). The orange and green dashed horizontal lines, for C3A and C3GC, respectively, 
enclose the best estimated CCSD(T) (panel a and b) and the final FN-DMC (panel c) interaction energies using the 
corresponding uncertainty estimates and stochastic error bars. 
The FN-DMC error bars indicate a stochastic error of 1-$\sigma$. 
The yellow bar denotes the minimum difference between CCSD(T) and FN-DMC ($\Delta_\text{min}$).
(a) CP-corrected and uncorrected LNO-CCSD(T) interaction energies using the aug-cc-pV$X$Z basis sets, 
as well as CBS($X$,$X+1$) extrapolation. 
(b) Convergence of half CP-corrected LNO-CCSD(T)/CBS(Q,5) interaction energies using a series of LNO thresholds as well as
Normal--Tight (N-T) and Tight--very Tight (T-vT) extrapolations.
(c) FN-DMC interaction energies with two nodal surfaces \yas{for C3GC} from DFT (PBE0 and LDA) and different time-steps (given in a.u.) \yas{for C3A and C3GC}. (d) C3A complex. (e) C3GC complex.
}
\end{figure*}

CCSD(T) and FN-DMC interaction energies involve 
multiple approximations.
Known sources of error to consider in our FN-DMC calculations are:
\begin{itemize}
    \item The fixed-node approximation which restricts the nodal-structure to that of the input guiding wavefunction.
    \item Time-step bias from the discretization of imaginary time for propagating the wavefunction.
    \item Pseudopotentials to approximate core electrons for each atom.
    \item Non-uniform quality of optimized trial wavefunctions for fragments and bound complex \yas{at larger time-steps}.
\end{itemize}
In obtaining CCSD(T) interaction energies, the sources of error are:
\begin{itemize}
    \item Local approximations of long-range electron correlation according to the LNO scheme.
    \item Single-particle basis representation of the CCSD(T) wavefunction.
    \item Neglected core electron correlation.
    \item Missing high-order many-electron contributions beyond CCSD(T).
\end{itemize}
In Fig.~\ref{fig3:convergence} we 
analyse the most critical approximations for each method on the example of the C3A and C3GC complexes, and we also consider the other remaining known sources of error in Methods. 

For the single-particle basis representation in CCSD(T) we employed conventional
correlation-consistent basis sets augmented with diffuse functions~\cite{AUGPVXZ}, aug-cc-pV$X$Z ($X$=T, Q, and 5) as shown in panel a) of Fig.~\ref{fig3:convergence}. 
The remaining BSI is alleviated using
extrapolation~\cite{CorrConv} toward the complete basis set (CBS) limit [CBS($X$,$X+1$), $X$=T, Q], and counterpoise (CP) corrections~\cite{BSSE}.
The local errors decrease systematically as the LNO threshold sets are tightened (Normal, Tight, very Tight) enabling 
extrapolations, \textit{e.g.} Normal--Tight (N--T), to estimate the canonical CCSD(T) interaction energy~\cite{LocalCC4} (see panel b) of Fig.~\ref{fig3:convergence}).
Exploiting the systematic convergence properties,
an upper bound for both the local and the BSI errors
can be given~\cite{cc-error-estimate}.

\pn{
Benchmarks presented previously for energy differences of a broad variety of systems showed  excellent overall accuracy
at the Normal--Tight extrapolated
LNO-CCSD(T)/CBS(T,Q) level (M1)~\cite{LocalCC4}. 
However, the BSI error bar of  1.0~kcal~mol$^{-1}$
and the local error bar of 2.2~kcal~mol$^{-1}$ obtained for C3GC at this M1 level are
impractical for a definitive comparison with FN-DMC. 
The next steps along both series of approximations towards chemical accuracy, \textit{i.e.} the use of very Tight LNO thresholds and the aug-cc-pV5Z basis set (M2),
have been enabled by our recent method development efforts~\cite{LocalCC3,LocalCC4,MRCC}.
With these better converged interaction energies, the M2 level uncertainty estimates are up to a factor of three smaller than at the M1 level.
Explicitly,  0.7 (0.4) kcal mol$^{-1}$ local (BSI) error 
estimate is obtained for C3GC. The same 
measures are the largest for C$_{60}$@[6]CPPA at the M2 level being 1.1 and 0.6 kcal mol$^{-1}$, respectively.}
Moreover, for the remaining L7 complexes, the local (BSI) uncertainty estimates 
indicate even better convergence of 0.1-0.4 (0.1-0.3) kcal mol$^{-1}$. 
Additional details are provided in Methods and in Section \ref{SMCC} of the Supplemental Material (SM).

FN-DMC has the advantage that the wavefunction is sampled in real-space without the need for the numerical 
representation of many-particle basis states thus \yas{reducing} sensitivity to the single-particle basis set.
Instead, pertinent sources of error in FN-DMC which we assess in Fig.~\ref{fig3:convergence} are the effects of the fixed-node approximation and the time-step bias. \pn{Note that these sources of error are not included in the FN-DMC stochastic error bars.} 

First, the different nodal surfaces from these DFT methods serve to indicate the dependence of the FN-DMC interaction energy on the nodal structure. Indeed, from Fig.~\ref{fig3:convergence}, we find no indication that the FN-DMC interaction energies of C3GC is affected by the nodal structure \pn{outside of the stochastic error bars}.
Second, FN-DMC energies are sensitive to the time-step and we rely on recent improvements in FN-DMC algorithms~\cite{zen2016boosting,Zen2019}, that enable convergence of time-steps as large as 0.05~a.u.
We used 0.03~a.u. and 0.01~a.u. time-steps to compute the interaction energies of C3A and C3GC. 
Fig.~\ref{fig3:convergence} indicates that the interaction energy is unchanged for C3A and C3GC (within the stochastic error) for the different time-steps considered here.
The time-step and fixed-node approximations perform similarly well for the coronene dimer and the buckyball-ring complex (see Section~\ref{SMDMC} of the SM).

\subsection*{Open challenges for next generation of many-body methods}

\yas{CCSD(T) and FN-DMC have been shown to agree with sub-chemical accuracy for small organic dimers~\cite{A24set2,Dubecky2016,Al-Hamdani2019}, molecular crystals~\cite{Zen2018}, and small physisorbed molecules on surfaces~\cite{waterongraphene,lih_dmc_cc}. Indeed, we also find good agreement in the absolute interaction energies for five of the eight complexes considered here.}
However, we find that the disagreement by several kcal mol$^{-1}$ in C$_{60}$@[6]CPPA particularly, cannot be explained by the controllable sources of error.
While both methods are highly sophisticated, they are still approximations to the exact solution of the many-electron Schr{\"o}dinger equation. 
Moreover, there can be non-trivial coupling between approximations within each method, which remain poorly understood for complex many-electron wavefunctions.
\pn{Here, we estimate the magnitude of additional approximations which are generally regarded as even more robust and contemplate potential strategies for improvements.}

\subsubsection{Are we there yet with FN-DMC?} 

\yas{The reported interaction energies of C2C2PD, C3GC, and C$_{60}$@[6]CPPA indicate that FN-DMC stabilizes the interacting complexes more weakly than CCSD(T). Therefore, one possibility for the discrepancy between the methods is that FN-DMC (as applied here) does not capture the correlation energy in the bound complexes sufficiently. Reasons for this can include the fixed-node approximation and more generally, insufficient flexibility in the wavefunction ansatz.

The Slater-Jastrow ansatz was applied here using a single determinant combined with a Jastrow factor containing explicit parameterizable functions to describe electron-electron, electron-nucleus, and electron-electron-nucleus interactions. We have evaluated FN-DMC interaction energies for different nodal structures for C3GC, C2C2PD, C$_{60}$@[6]CPPA and in all cases the FN-DMC interaction energies are in 1-$\sigma$ agreement (see Section~\ref{SMDMC} of SM) with stochastic errors that are mostly under 1 kcal mol$^{-1}$. Among these systems, the largest potential deviation ($\Delta_\text{max}$) due to the fixed-node error is estimated to be $\sim$3.7 kcal mol$^{-1}$ in C$_{60}$@[6]CPPA. Although this potentially large source of error is not enough to explain the 8.3 kcal mol$^{-1}$ $\Delta_\text{min}$ disagreement with CCSD(T), it remains a pertinent issue for establishing chemical accuracy. Reducing the fixed-node error, for example by using more than one Slater determinant to systematically improve the nodal structure, in such large molecules remains challenging~\cite{Morales2012,Scemama2016}. Promising alternatives include the Jastrow antisymmetrized geminal power approach which has recently been shown to recover near-exact results for a small, strongly correlated cluster of hydrogen atoms~\cite{Genovese2019}. 

The Jastrow factor is a convenient approach to increase the efficiency of FN-DMC since in the zero time-step limit and with sufficient sampling, the FN-DMC energy is independent of this term. However, the quality of the Jastrow factor can be non-uniform for the bound complex and the non-interacting fragments, which can introduce a bias at larger time-steps. The recent DLA method in FN-DMC reduces this effect~\cite{Zen2019} and was applied to the C$_{60}$@[6]CPPA complex reported in Table \ref{tab:my-table} and also tested for GGG, C3A, and C2C2PD complexes (see Methods for further details). In all cases, FN-DMC with DLA is in decent agreement (within 1-$\sigma$) with non-DLA FN-DMC interaction energies. For example, the C2C2PD FN-DMC interaction energy with DLA is $-17.4 \pm 0.5$ kcal mol$^{-1}$ whilst with standard LA, it is $-18.1 \pm 0.4$ kcal mol$^{-1}$. Moreover, the interaction strengths tend towards being weaker with DLA in the systems we consider,\textit{ i.e.} further from the CCSD(T) interaction energies. As such, the discrepancy between FN-DMC and CCSD(T) remains regardless of any potential error from the Jastrow factor in our findings. 

We estimate the error from the use of Trail and Needs pseudopotentials~\cite{Trail2005,Trail2005b} in FN-DMC
at the Hartree-Fock (HF) level using interaction energy of C2C2PD. We find 0.1 kcal mol$^{-1}$ difference in the HF interaction energy with the employed pseudopotentials and without (\textit{i.e.} all-electron) which is well within the acceptable uncertainty for our findings. In addition, Zen \textit{et al.}~\cite{Zen2018} have previously compared Trail and Needs pseudopotentials with correlated electron pseudopotentials~\cite{Trail2013} at the FN-DMC level using the binding energy of an ammonia dimer and found agreement within 0.1 kcal mol$^{-1}$. 

In principle, a more flexible wavefunction ansatz allows a more accurate many-body wavefunction to be reached in DMC, thus recovering electron correlation more effectively. To this end, recently introduced machine learning approaches~\cite{NNsolvSchrodingerEq,NNsolvSchrodingerEq2} are promising but more expensive due to the considerable increase in parameters. However once feasible, a systematic assessment of the amount of electron correlation recovered by these different ansatze in non-covalently bound systems will bring valuable insight to the current puzzle.

}

\subsubsection{Potential avenues for improvement upon CCSD(T)}

\pn{Considering the complexes exhibiting significant $\pi$-$\pi$ interactions, 
CCSD(T) is found to predict stronger interaction than FN-DMC.}
As some of the individually negligible but collectively important long-range interactions
are estimated in local CC methods, 
\pn{these potentially overestimated interaction energy contributions could benefit from a higher-level theoretical description~\cite{fragmentbook,PNOCCreview}.}
In the case of the LNO scheme, the majority of the local approximations have marginal effect on the interaction energies when very Tight settings are employed~\cite{LocalCC4}. 
\pn{For instance, long-range interactions that do not benefit from the full CCSD(T) treatment add up to at most 2.9 kcal mol$^{-1}$ for the interaction energy of C$_{60}$@[6]CPPA. The presented error estimate allowing for almost 40\% error in this term reliably covers this approximation.}
While these and other non-negligible LNO approximations are closely monitored (see Section \ref{localconvg} of SM),
remaining uncertainties outside of the presented error bars cannot be ruled out. 
All in all, the convergence measures assessing the local errors of LNO-CCSD(T) interaction energies indicate at least 97.4\% or 1.1 kcal mol$^{-1}$ certainty.

The employed single-particle basis sets perform exceptionally well for CCSD(T) computations of small molecules~\cite{AUGPVXZ,CorrConv}, 
but approaching the CBS limit of CCSD(T) for large systems is mostly an uncharted territory in the literature~\cite{LocalCC4,PNOCCreview}.
\pn{The agreement of CP corrected CBS(T,Q), CBS(Q,5), and uncorrected CBS(Q,5) within 0.06--0.36 kcal mol$^{-1}$ is highly satisfactory
(see Sect. \ref{CCbasis} of SM). Furthermore, the CBS(5,6) results obtained with the aug-cc-pV6Z basis set for GGG are 
fully consistent with the  CBS(Q,5) interaction energies (see Sect. \ref{CCbasis} of SM).}
CC methods exploiting explicitly correlated wavefunction forms~\cite{PNOCCreview,DLPNO-CCSD(T)-F12} as well as 
alternatives to the conventional Gaussian basis sets~\cite{QMC_CC_solids,CCSD_DMC_solidH,MRA_CC2} have emerged recently,
which could provide independent verification of the systematic convergence studies performed here.

The higher-order contribution of three-, four-, etc. electron processes \pn{on top of CCSD(T)}
are usually found to be negligible for weakly-correlated molecules~\cite{A24set2}. 
However, the available numerical experience is limited to complexes below about a dozen atoms, 
and for some highly polarizable systems the beyond CCSD(T) treatment of three-electron processes 
has been shown to contribute significantly to three-body dispersion~\cite{LR3bodydispersion}. 
\pn{The weakly-correlated nature of all complexes is indicated by 
that the perturbative (T) contribution to the total correlation energy component of the CCSD(T) interaction energy 
is consistently around 18--20\%. 
Additionally, the CC amplitude based measures all point to pure dynamic correlation. 
According to our LNO approximated estimations for the GGG complex,
the infinite-order three-electron terms on top of the perturbative treatment of (T) is only about -0.01~kcal~mol$^{-1}$, while 
the perturbative four-electron contribution~\cite{Pert} is around -0.02~kcal~mol$^{-1}$ (see Sect. \ref{coreTQ} of SM).
Due to the extreme computational cost of such computations,}
it remains an open and considerable challenge to establish, \pn{on a representative sample size,} 
 whether the contribution of 
higher-order processes is within sub-chemical accuracy for larger and more complex molecules.

\pn{
The effect of core correlation is expected to be very small, thus we attempted to evaluate it independently from the numerical noise of the 
other approximations. For instance, the all electron interaction energy of C2C2PD is even stronger than the frozen core one at second-order by 0.2 kcal mol$^{-1}$ (4.6~cal~mol$^{-1}$ per C atom).
All in all, core and higher-order correlation effects appear to strengthen the  CCSD(T) interaction energies  and slightly increase the deviation compared to FN-DMC.}

\subsubsection{Insights from experiments and density-functional approximations}
Experimental binding energies or association constants of supramolecular complexes are particularly valuable, when available, but also have their limitations as back-corrections are needed to separate the effects of thermal fluctuations and solvent effects for example~\cite{dissocEreview16}. In the case of C$_{60}$@[6]CPPA for example, the association constant is measured in a benzene solution and indicates a stable encapsulated complex, but one which could not be well-characterised by X-ray crystallography; purportedly due to the rapid rotation of the buckyball guest~\cite{Kawase2003}. Instead, a non-fully encapsulated structure was successfully characterized using toluene anchors on the buckyball. This demonstrates a number of physical leaps that exist between what can be measured and what can be accurately computed.  

Other high-level methods, such as the full configuration interaction quantum Monte-Carlo (FCI-QMC) method~\cite{QMC_CC_solids,CCSD_DMC_solidH},
can be key to assessing the shortcomings from major approximations such as the fixed node approximation and static correlation. Once  the severe scaling with system size associated with FCI-QMC and similar methods is addressed, larger molecules will become feasible. However, in the present time the lack of references in large systems remains a salient problem.  

\begin{figure}[htb]
\includegraphics[width=8.5cm]{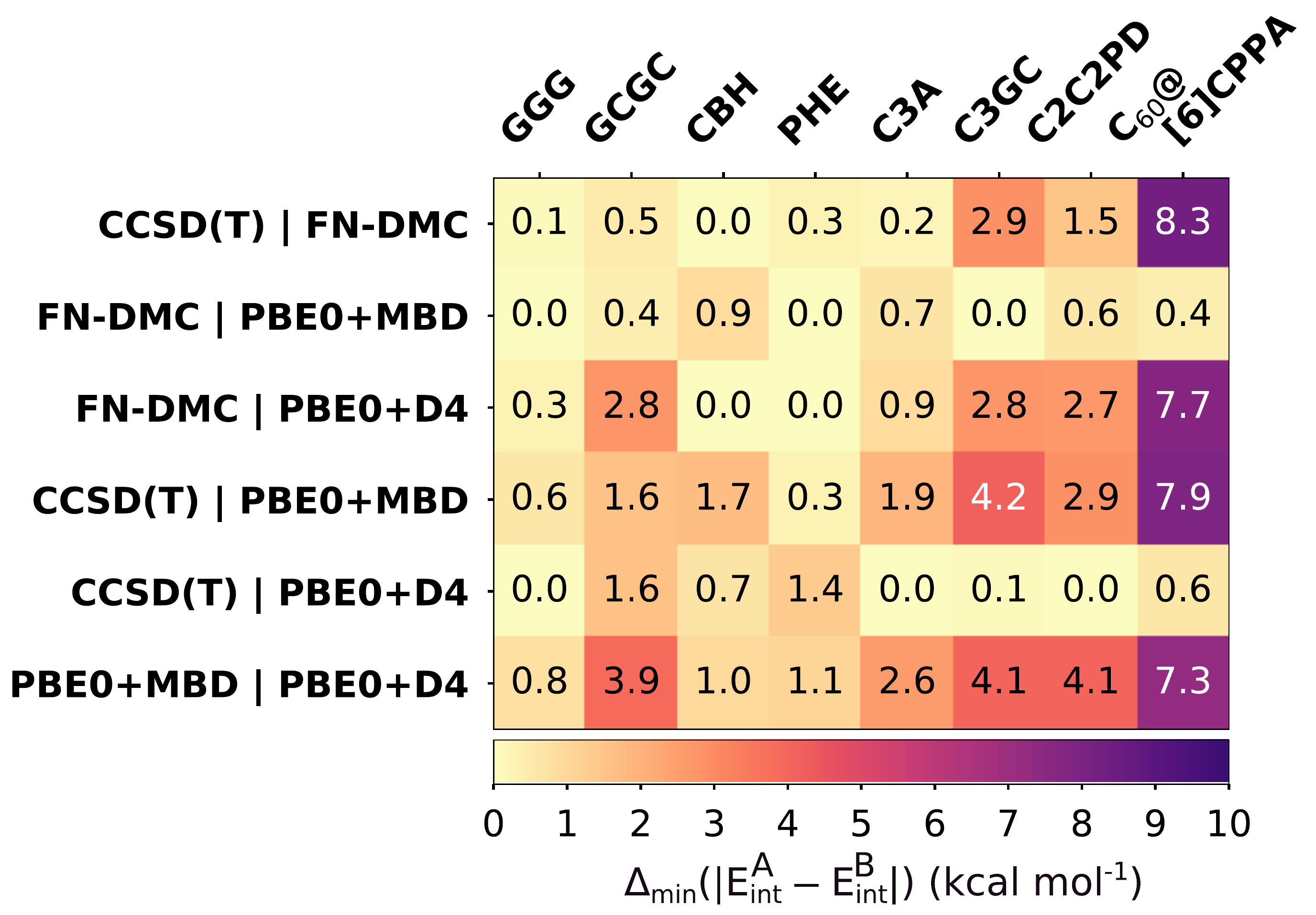}
\caption{\label{fig4:dftcomparison} 
$\Delta_{\text{min}}(|E^{\text{A}}_{\text{int}} - E^{\text{B}}_{\text{int}}|)$ is shown which is the smallest absolute difference in $E_{\text{int}}$ between pairs of methods ($|E^{\text{A}}_{\text{int}} - E^{\text{B}}_{\text{int}}|$) in kcal mol$^{-1}$, with methods $\text{A}$ and $\text{B}$ indicated along the vertical axis. $\Delta_{\text{min}}$ takes into account the error estimates for CCSD(T) and FN-DMC to show smallest possible differences with respect to these reference methods. The DFT methods have no quantified uncertainty estimates associated with them. The compared methods are: CCSD(T), FN-DMC, PBE0+MBD and PBE0+D4. The supramolecular complexes are those in the L7 data set and the C$_{60}$@[6]CPPA buckyball-ring complex.}
\end{figure}

The scarcity of reference information has an impact on all other modelling methods, \pn{including density-functional approximations (DFAs), semi-empirical,  force field or machine learning based models, etc.} 
 which are validated  \pn{or parameterized} based on higher-level benchmarks. 
In particular, there is a race to simulate larger, more anisotropic, and complex materials, accompanied by a \jgb{difficulty} of choice for modelling methods.
To demonstrate \pn{the consequences of inconsistent references}, Fig. \ref{fig4:dftcomparison} shows interaction energy discrepancies obtained with 
 DFAs, PBE0+D4~\cite{dftd4} and PBE0+MBD~\cite{MBD}, that are both designed to capture all orders of many-body dispersion interactions in different manner. 
Intriguingly, the PBE0+D4 method is in close agreement with CCSD(T) (mean absolute deviation, MAD=1.1~kcal~mol$^{-1}$), 
whereas PBE0+MBD is closer to FN-DMC (MAD=1.5~kcal~mol$^{-1}$)\pn{, but their performance is hard to characterize when  CCSD(T) and FN-DMC disagree.}
Moreover, we decomposed the interaction energies from the DFAs into dispersion components and find that, for C$_{60}$@[6]CPPA the main difference between PBE0+MBD and PBE0+D4 is 6.5~kcal~mol$^{-1}$ 
in the two-body dispersion contribution. Differences in beyond two-body dispersion interactions are smaller and at most 1.6~kcal~mol$^{-1}$ in C$_{60}$@[6]CPPA. 

\section*{Discussion}
Until now, disagreements between reference \pn{interaction energies of extended organic complexes} have typically been ascribed to \pn{unconverged results due to} practical bottlenecks. Here, we report highly-converged results at the frontier of wavefunction based methods; uncovering a disconcerting level of disagreement in the interaction energy for three supramolecular complexes. We have computed interaction energies from CCSD(T) and FN-DMC for a set of supramolecular complexes of up to 132 atoms exhibiting challenging intermolecular interactions. The accuracy of these methods have been repeatedly corroborated in the domain of dozen-atom systems 
\pn{with single-reference character} and here we find CCSD(T) and FN-DMC are in excellent agreement for five of the supramolecular complexes suggesting that these methods are able to maintain remarkable accuracy in some larger molecules.
However, FN-DMC and CCSD(T) interaction energies disagree by 1.5 kcal mol$^{-1}$ in the coronene dimer (C2C2PD), 2.9 kcal mol$^{-1}$ in GC base pair on circumcoronene (C3GC) and 8.3 kcal mol$^{-1}$ in a buckyball-ring complex (C$_{60}$@[6]CPPA).
These disagreements are cemented by reporting sub-kcal mol$^{-1}$ stochastic errors in FN-DMC and a systematically converging series of local CCSD(T) interaction energies accompanied by uncertainty estimates approaching chemical accuracy.
Therefore, despite our best efforts to suppress all controllable sources of error, the marked disagreement of FN-DMC and CCSD(T) prevents us from providing conclusive reference interaction energies for these three complexes. 
Such large differences in interaction energies surpass the widely-sought 1 kcal mol$^{-1}$ chemical accuracy and indicate that the highest level of caution is required even for our most advanced tools when employed at the hundred-atom scale.

The supramolecular complexes we report feature $\pi-\pi$ stacking, hydrogen bonding, and intermolecular confinement, that are ubiquitous across natural and synthetic materials.
Thus our immediate goals are to elucidate the sources of the underlying discrepancies and to
explore the scope of systems where such deviations between reference wavefunction methods occur.
\pn{Well-defined reference interaction energies and the better characterization of their predictive power have growing importance
as they} \jgb{are frequently applied} 
\pn{in chemistry, material, and biosciences.}
Our findings should motivate cooperative efforts between experts of computational and experimental methods in obtaining well-defined interaction energies 
and thereby extending the predictive power of first principles approaches across the board.

\section*{Methods}

The L7 structures have been defined by Sedlak \textit{et al.}~\cite{L7set} and structures can be found on the begdb database~\cite{Rezac2008}.
Note that the interaction energy, $E_{\text{int}}$, is defined with respect to two fragments even where the complex consists of more than two molecules (as in GGG, GCGC, PHE, and C3GC):
\begin{equation} \label{IEdef}
    E_{\text{int}} = E_{\text{com}} - E_{\text{frag}}^1 - E_{\text{frag}}^2
\end{equation}
where $E_{\text{com}}$ is the total energy of the full complex, and $E_{\text{frag}}^1$ and $E_{\text{frag}}^2$ are the total energies of isolated fragments 1 and 2, respectively. The fragment molecules have the same geometry as in the full complex, \textit{i.e.} not relaxed. Further details on the configurations can be found in the SM and in Ref.~\citenum{L7set}.

The C$_{60}$@[6]CPPA complex is based on similar complexes in previous theoretical and experimental works~\cite{c60_at_CPPs_angew,s8_chemcomm,Hermann2017} and has been chosen to represent confined $\pi$-$\pi$ interaction that are numerically still tractable by our methodologies. Its geometry has been symmetrized to $D_{3d}$ point group, the individual fragments of C$_{60}$ and [6]CPPA are kept frozen ($I_h$ and $D_{6h}$, respectively). 
The structure is provided in the SM. 

\subsection*{The local natural orbital CCSD(T) method}

In order to reduce the $N^7$-scaling of canonical CCSD(T) with respect to the system size ($N$), 
the inverse sixth power decay
of pairwise interactions can be exploited (local approximations) 
and the wavefunction can be compressed further via natural orbital (NO) techniques.~\cite{fragmentbook} 
Building on such cost-reduction techniques a number of highly-efficient local CCSD(T) methods emerged in the past decade~\cite{fragmentbook,DLPNO-CCSD(T),DLPNO-CCSD(T)-F12,PNO-CCSDHattig,PNOCCreview,LocalCC3,LocalCC4}.
As the local approximation free
CCSD(T) energy can be approached by the simultaneous improvement of all local truncations in most of these 
techniques, in principle, all local CCSD(T) methods are expected to converge to the same interaction energy.
Here we employ the local natural orbital CCSD(T) [LNO-CCSD(T)] scheme~\cite{LaplaceT,LocalCC3}, which, for the studied systems, brings the feasibility of exceedingly 
well-converged CCSD(T) calculations in-line with FN-DMC. 
The approximations of the LNO scheme automatically adapt to the complexity of the underlying wavefunction 
and enable systematic convergence towards the exact CCSD(T) correlation energy, with up to
99.99\% accuracy using sufficiently tight settings~\cite{LocalCC4} .

The price of improvable accuracy is that
the computational requirements can drastically increase depending on the nature of the wavefunction: 
while LNO-CCSD(T) has been successfully employed for 
macromolecules, such as small proteins at the 1000 atom range~\cite{LocalCC3,LocalCC4}, sizable 
long-range interactions appearing in the here studied complexes pose a challenge for any local CCSD(T) method~\cite{PNOCCreview,DLPNO-CCSD(T),DLPNO-CCSD(T)-F12,LocalCC4}.
This motivated the implementation of several recent developments in our algorithm and computer code over the lifetime of this project,
which cumulatively resulted in about 2-3 orders of
magnitude decrease in the time-to-solution and data storage requirement of LNO-CCSD(T)~\cite{LaplaceT,LocalCC3,LocalCC4},
and made well-converged computations feasible for all complexes.
For instance, we have designed
a massively parallel conventional CCSD(T) code specifically for applications within the LNO scheme~\cite{MPICCSDpT} 
and integrated it with our highly optimized LNO-CCSD(T) algorithms~\cite{LaplaceT,LocalCC3,LocalCC4}.
Here, we report the first large-scale LNO-CCSD(T) applications which exploit the resulted high performance capabilities using the most recent
implementation of the {\sc Mrcc} package~\cite{MRCC} (release date February 22, 2020).

\subsection*{Computational details for CCSD(T)}

The LNO-CCSD(T)-based CCSD(T)/CBS estimates 
were obtained as the 
average of CP-corrected and uncorrected (``half CP'')~\cite{BSSE}, Tight--very Tight extrapolated LNO-CCSD(T)/CBS(Q,5) 
interaction energies~\cite{LocalCC4}. Except for C3A, C3GC, and C$_{60}$@[6]CPPA, the CBS(Q,5) notation refers to CBS extrapolation~\cite{CorrConv} using 
aug-cc-pV$X$Z basis sets~\cite{AUGPVXZ} with $X$=Q and 5. For C3A, C3GC, and C$_{60}$@[6]CPPA, a Normal 
LNO-CCSD(T)/CBS(Q,5)-based BSI correction ($\Delta_\text{BSI}$) was added to the
Tight--very Tight extrapolated LNO-CCSD(T)/aug-cc-pVTZ interaction energies, exploiting the parallel convergence of the 
LNO-CCSD(T) energies for these basis sets~\cite{LocalCC4}. 
Local error bars shown, \textit{e.g.} on panel b) of Fig. \ref{fig3:convergence}  
are obtained via the extrapolation scheme of Ref.~\cite{LocalCC4}. 
Error bars accompanying the LNO-CCSD(T) interaction energies of Fig. \ref{fig2:mainres} and Table \ref{tab:my-table}, and determining the interval enclosed by the dashed lines
on panels a) and b) of Fig. \ref{fig3:convergence}
are the sums of the BSI and local error estimates. The BSI error measure is the maximum of two separate error estimates: the difference between
CP-corrected and uncorrected CBS(Q,5) energies, and the difference between CP-CBS(T,Q) and CP-CBS(Q,5) results. 
This BSI error bar is increased with an additional term if $\Delta_\text{BSI}$ is employed according to Sect.~\ref{CCbasis} of the SM. 
The local error bar of the best estimated CCSD(T) results (see Table \ref{lccres}
of the SM) is obtained from the difference of the Tight and very Tight LNO-CCSD(T) results evaluated with the largest possible basis sets~\cite{LocalCC4}.

\subsection*{Computational details for FN-DMC}

Our FN-DMC calculations use the Slater-Jastrow ansatz with the single Slater determinants obtained from DFT. The Jastrow factor for each system contains explicit electron-electron, electron-nucleus, and \yas{three-body} electron-electron-nucleus terms. The parameters of the Jastrow factor were optimized for each complex using the variational Monte Carlo (VMC) method and the varmin algorithm which allows for systematic improvement of the trial wavefunction, as implemented in CASINO v2.13.610~\cite{Needs_2009}. Note that bound complexes were used in the VMC optimizations and the resulting Jastrow factor was used to compute the corresponding fragments.  
\yas{All systems were treated in real-space as non-periodic open systems in VMC and FN-DMC.}

We performed FN-DMC with the locality approximation (LA) for the non-local pseudopotentials~\cite{Mitas1991} and 0.03 a.u. time-step for all L7 complexes. Smaller time-steps of 0.003 and 0.01 a.u. were also used to compute the interaction energy of the C2C2PD complex and the interaction energy was found to be in agreement within the stochastic error bars with all three time-steps. 

The C$_{60}$@[6]CPPA complex exhibited numerical instability \yas{using the standard LA}. This prevented sufficient statistical sampling and therefore we computed this complex with two alternative and more numerically stable approaches. First, the energy reported in Fig.~\ref{fig2:mainres} and Table~\ref{tab:my-table} is using the recently developed determinant localization approximation (DLA)~\cite{Zen2019} implemented \yas{in CASINO v2.13.809}~\cite{Needs_2009}. The DLA gives: (i) better numerical stability \yas{than the LA algorithm} allowing for more statistics to be accumulated, \yas{(ii)} smaller dependence on the Jastrow factor, and \yas{(iii)} addresses an indirect issue related to the use of \yas{non-local} pseudopotentials. Second, the T-move approximation~\cite{Casula2006} (without DLA) \yas{was also applied to C$_{60}$@[6]CPPA for comparison}. \yas{The T-move scheme is more numerically stable than the standard LA algorithm but is also more} time-step dependent and therefore we used results from 0.01 and 0.02 a.u. time-steps to extrapolate the interaction energy to the zero time-step limit, as reported in SM. The extrapolated interaction energy \yas{with the T-move scheme} is $-31.14 \pm 2.57$ kcal mol$^{-1}$ using LDA nodal structure and $-29.16 \pm 2.33$ kcal mol$^{-1}$ using PBE0 nodal structure. Due to the large stochastic error on these results, we report the better converged DLA-based interaction energy (with PBE0 nodal structure) in the main results, but we note that all three predictions from FN-DMC agree within the statistical error bars. \yas{Furthermore, as the DLA is less sensitive to the Jastrow factor at finite time-steps, we have also tested the interaction energies of GGG, C3A, and C2C2PD complexes, finding agreement with the LA-based FN-DMC results within one standard deviation. Further details can be found in the SM.}

The initial DFT orbitals (which define the nodal structure in FN-DMC) were prepared using PWSCF in Quantum Espresso v.6.1~\cite{Giannozzi2009}. Trail and Needs pseudopotentials~\cite{Trail2005,Trail2005b} were used for all elements, with a plane-wave energy cut-off of 500 Ry. 
The plane-wave representation of the molecular orbitals from PWSCF were expanded in terms of B-splines. Since PWSCF uses periodic boundary conditions, all complexes were centered in an orthorhombic unit cell with a vacuum spacing of $\sim8$~\AA~in each Cartesian direction \yas{to ensure that the single-particle orbitals are fully enclosed}. LDA orbitals were used for L7 complexes and in addition, PBE0 orbitals were also considered for C2C2PD, C3GC, and C$_{60}$@[6]CPPA. In all cases, the final FN-DMC interaction energy from LDA and PBE0 nodal structures are in agreement within the stochastic errors. 

\section*{Acknowledgments}
We thank Dr. Andrea Zen for discussions. We thank HPC staff for their support and access to the IRIS cluster at the University of Luxembourg and to the DECI resource Saga based in Norway at Trondheim with support from the PRACE aisbl (NN9914K). \textbf{Funding:}
YSA thanks funding from NIH grant number R01GM118697 and is supported by The National Centre of Competence in Research (NCCR) Materials Revolution: Computational Design and Discovery of Novel Materials (MARVEL) of the Swiss National Science Foundation (SNSF). 
The work of PRN is supported by the \'UNKP-19-4-BME-418 New National Excellence Program of the Ministry
for Innovation and Technology and the J\'anos Bolyai Research Scholarship of the Hungarian Academy of Sciences.
JGB acknowledges support from the Alexander von Humboldt foundation.
MK is grateful for the financial support from the National Research,
Development, and Innovation Office (NKFIH, Grant No. KKP126451) and 
the BME-Biotechnology FIKP grant of EMMI (BME FIKP-BIO).
AT acknowledges financial support from the European Research Council (ERC-CoG grant BeStMo). 

\section*{Author contributions} 
YSA and PRN contributed equally to this work. Major \textit{investigation} was conducted by PRN (performing coupled cluster calculations) and YSA (performing quantum Monte Carlo calculations). JGB performed PBE0+D4 calculations and supporting validation calculations. DB performed PBE0+MBD calculations. The work has been \textit{conceptualized} by YSA and AT with additional contribution from JGB and PRN. \textit{Software development} to expand the application of LNO-CCSD(T) method in this work has been conducted by PRN and MK. PRN performed \textit{formal analysis} to obtain uncertainty estimates from LNO-CCSD(T) data. The \textit{original draft} of the manuscript was written by YSA and PRN. Additional \textit{review and editing} of the manuscript was undertaken by JGB, AT, and MK. \textit{Project administration} was led by YSA with contribution from JGB and AT. JGB and AT supervised the work.   


{\centering
\section*{Supplemental Material}
}

\tableofcontents

\section{Details of CCSD(T) computations} \label{SMCC}

Definitions:
\begin{itemize}
    \item Interaction energy: according to the Methods Section of the main text, the difference of the complex's energy consisting of all molecules and of the two subsystem energies, using unrelaxed structures for the latter. Notation: 
    $E^{\mathrm{LNO-CCSD(T)}}_\mathrm{Y}$[aug-cc-pVXZ], where $Y$ refers to the level of local approximations ({\it Normal, Tight}, or {\it very\,Tight}) and X labels the cardinal number of the basis set.  
    \item counterpoise (CP) corrected interaction energy: the energy of the subsystems are evaluated for the interaction energy expression using all single-particle basis functions of the complete complex including basis functions residing on the atomic positions of the other subsystem.  
    \item local error bar: difference of the Tight and very Tight LNO-CCSD(T) results evaluated with the largest possible basis set.
    \item basis set incompleteness (BSI) error bar:  
    maximum of two BSI error indicators, which are the difference of the
CP corrected and uncorrected LNO-CCSD(T)/CBS(Q,5) interaction energies, and the difference of CP corrected 
 LNO-CCSD(T)/CBS(T,Q) and LNO-CCSD(T)/CBS(Q,5) interaction energies.
\end{itemize}

\subsection{Convergence of local approximations} \label{localconvg}

\pn{ 
The LNO-CCSD(T) energy expression reformulates the CCSD(T) energy in terms of localized molecular orbitals (LMOs, $\ip,\jp$)~\cite{LocalCC2,LaplaceT,LocalCC3}:
\begin{equation}\label{lnoccsdptE}
E^{\mathrm{LNO-CCSD(T)}}=  \sum_\ip \left [  \delta E_{\ip}^\mathrm{CCSD(T)} + \Delta E_{\ip}^\mathrm{MP2} +
\frac12 \sum\limits_{\jp}^\text{ distant} \delta E_{\ip\jp}  \right ].
\end{equation}
The correlation energy contribution of distant LMO pairs is obtained at the level of approximate MP2~\cite{LocalMP2,LocalCC3} (third term), while
all remaining LMO-pairs contribute to the CCSD(T) level treatment (first term). 
For the latter, first, local natural orbitals (LNOs) are constructed individually for each LMO at the MP2 level using 
a large domain of atomic and correlating (virtual) orbitals surrounding the LMO. The 
$ \delta E_{\ip}^\mathrm{CCSD(T)} $ contribution is then computed in this compressed LNO orbital space, while the second term 
of Eq. (\ref{lnoccsdptE}) represents a correction for the truncation of the LNO space at the MP2 level of theory.
}

The convergence of all approximations in LNO-CCSD(T) can be assessed  via the use of 
pre-defined threshold sets, which provide systematic improvement simultaneously for all approximations of the LNO scheme~\cite{LocalCC,LocalCC2,LocaldRPA,LocalMP2,LaplaceT,LocalCC3,LocalCC4}. 
In this series of threshold sets ({\it Normal}, {\it Tight}, {\it very\,Tight}),
the accuracy determining cutoff parameters are tightened in an exponential manner~\cite{LocalCC4}. For instance, the  {\it very\,Tight} set collects 
an order of magnitude tighter truncation thresholds than those of the {\it Normal} set, which is the default choice.
The convergence behavior of the LNO-CCSD(T) interaction energies separates the studied complexes (see Fig. 2 of manuscript) into two groups. 
For GGG, PHE, CBH, and GCGC we observe rapid convergence toward the corresponding canonical CCSD(T) interaction
energy as indicated, \textit{e.g.} by the local error estimates collected in Table~\ref{lccres}.
The excellent convergence is apparent as the differences of the {\it Tight} and {\it very\,Tight} interaction energies 
are all in the 0.1-0.3 kcal mol$^{-1}$ range for these four complexes. This uncertainty range is highly satisfactory
for the local approximations considering that the estimated
basis set incompleteness (BSI) errors for LNO-CCSD(T) are also comparable.
\begin{table}[H]
\footnotesize
\caption{Best converged [{\it Tight}--{\it very\,Tight} extrapolated LNO-CCSD(T)/CBS(Q,5) based] CCSD(T) interaction energies (IEs) 
and corresponding error estimates 
with full, half, and no CP correction. Our best estimates are highlighted in bold and are used throughout the manuscript.}
\label{lccres}
\begin{center}
\begin{tabular}{l rrrrrrrr}\hline\hline
System:           & \;\; GGG \;\; & \;\;CBH \;\; & \;\;GCGC \;\; & \;\;C3A \;\;	 & \;\;C2C2PD \;\;& \;\;PHE \;\;		& \;\;C3GC \;\;		& \;\;C$_{60}$@[6]CPPA  \;\;  \\ \hline
IE, no CP               & -1.98&    -10.93 &    -13.38&     -16.79	&    -20.36 	&    -25.34	&    -28.67	&    -41.60 \\
IE, CP                  & -2.20&    -11.10 &    -13.80&     -16.28	&    -20.84 	&    -25.38	&    -28.73	&    -41.89 \\
IE, half CP       & \bf   -2.09&\bf -11.01 &\bf -13.59 &\bf -16.53	&\bf -20.60 	&\bf -25.36	&\bf -28.70	&\bf -41.74 \\ \hline
Local error             & 0.09 &0.10		&0.16 &0.42	&0.38	&0.07		&0.65	&1.10 \\
BSI error               & 0.11 &0.06		&0.22 &0.24	&0.24	&0.12		&0.19	&0.36 \\
$\Delta_\mathrm{BSI}$ error &	&	  &     & 0.10 &	&	&0.17	&0.25  \\
Total error             & 0.20 &0.15	&0.39	&0.75 &0.62	&0.18		&1.01	&1.71 \\
 \hline \hline
\end{tabular}
\end{center}
\end{table}

\begin{figure}[ht!]
\includegraphics[width=0.9\textwidth]{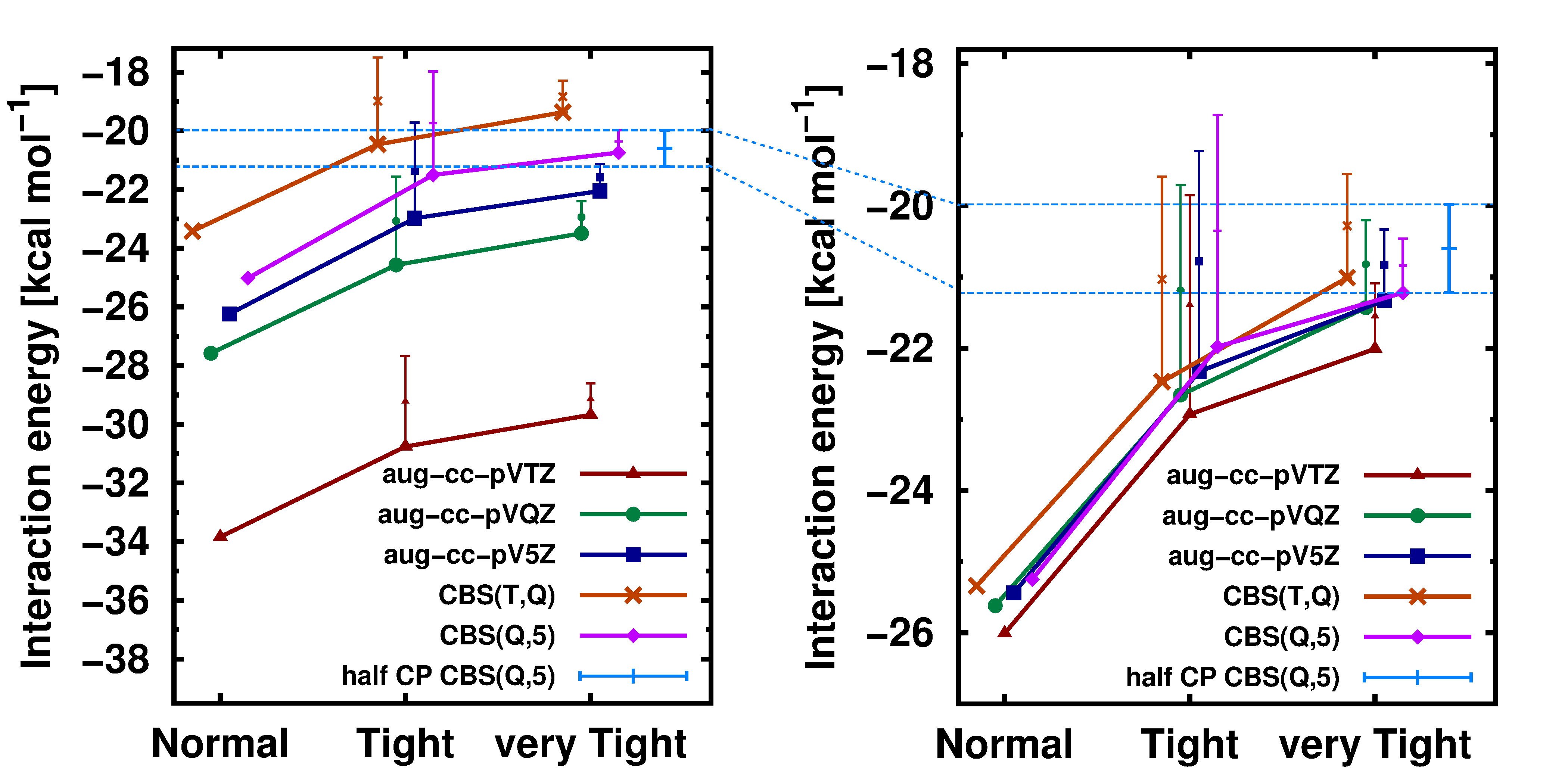}
  \caption{Convergence of LNO-CCSD(T) interaction energies for the C2C2PD complex with various basis sets and
LNO threshold sets. The left (right) panel collects results obtained without (with) CP correction.  The {\it Normal}--{\it Tight} and the {\it Tight}--{\it very\,Tight} 
extrapolated results are plotted with a smaller point size also at the 
{\it Tight} and {\it very\,Tight} x axis labels, respectively, and 
they are accompanied by error bars indicating the uncertainty estimate of the local 
approximations at that level.
For comparison the best CCSD(T)/CBS estimate 
[Tight-very Tight approximated, half CP corrected LNO-CCSD(T)/CBS(Q,5)] result and its corresponding uncertainty estimate is depicted on both panels via the light blue
error bars and dashed horizontal lines.
Note the different y ranges of the two panels as highlighted by dashed blue lines connecting the two panels.  
Also note that symbols corresponding to a given basis set are slightly shifted along the x-axis to improve visibility for all data points.}
  \label{coronene_basis}
\end{figure}

Consequently, we perform an even more thorough analysis of the local errors for the remaining four complexes,  
C2C2PD, C3A, C3GC, and C$_{60}$@[6]CPPA, where the 
local error estimate of the
LNO-CCSD(T) interaction energies is larger than 
0.3
kcal mol$^{-1}$.  
The convergence patter with {\it Normal}, {\it Tight}, and {\it very\,Tight} settings of the C3A and C3GC interaction energies 
is shown on the panel b) of Fig. \ref{fig3:convergence} of the main text. The monitored convergence is monotonic
and the remaining local error is about halved in each step, 
as observed for multiple systems previously~\cite{LocalCC4} as well for the above four complexes. 
Additionally, the {\it Normal}--{\it Tight} and the {\it Tight}--{\it very\,Tight} based CCSD(T) estimates 
(data points with error 
bars on panel b) of Fig.~\ref{fig3:convergence} of the main text) agree closely, 
and the {\it Tight}--{\it very\,Tight} error bars are enveloped by the 
{\it Normal}--{\it Tight}  ones. 
The same trends can be observed in Fig.~\ref{coronene_basis} for the coronene dimer, where LNO-CCSD(T) interaction energies 
are collected with all investigated basis sets and all three LNO threshold combinations. Again, 
the convergence patterns with the improving local approximations are  parallel
for all basis sets, the {\it Normal}--{\it Tight} and {\it Tight}--{\it very\,Tight} estimates agree within 0.5~kcal~mol$^{-1}$, and 
the {\it Tight}--{\it very\,Tight} error bars are  2-3 times narrower. 
Although, in the case of  C$_{60}$@[6]CPPA, the {\it Normal} to {\it very Tight} series is 
only available with the aug-cc-pVTZ basis set, the 0.2 kcal mol$^{-1}$ agreement of 
{\it Normal}--{\it Tight} and the {\it Tight}--{\it very\,Tight} based CCSD(T) estimates and the threefold
improvement provided by the {\it Tight}--{\it very Tight} error bar over the {\it Normal}--{\it Tight} one
illustrate  analogous behavior to the cases of C2C2PD, C3A, and C3GC.

\pn{
Considering briefly the corresponding absolute energies collected in Sect. \ref{absenergies}, 
one can observe that the LNO-CCSD(T) correlation energies are also sufficiently well converged. 
For instance, for the C3GC complex {\it very Tight} based results provide four (almost five) converged significant digits in the 
correlation energies, which is then reliably translates into the observed cca. 0.65 kcal mol$^{-1}$ local 
uncertainty of the LNO-CCSD(T) interaction energies. }

\pn{ 
One can also consider internal convergence indicators besides the total energy. 
At the very Tight level, the $\pi$--$\pi$, $\pi$--$\sigma$, and also the majority of the $\sigma$--$\sigma$ orbital
 interactions benefit from the full CCSD(T) treatment for all complexes.
Additionally, none of the remaining weak electronic interactions, 
contributing only about 0.01\% or lower portion of the correlation energy,
are neglected, they are, however, approximated via second-order pair energies~\cite{LocalCC3,LocalCC4}.
At the very Tight level, the orbital domains employed for the LNO-CCSD(T) treatment include all atoms, all atomic orbitals, and the majority of the 
correlating (virtual) space, spanned by, on the average, 80--95\% of the orbitals of the entire complexes.}

\subsection{Single-particle basis set convergence} \label{CCbasis}

Regarding the convergence of the interaction energies with respect to the single particle basis set,
we rely on approaches used routinely in wavefunction computations on small  molecules.
 Dunning's correlation consistent basis sets~\cite{AUGPVXZ} employed here are designed to 
systematically approach the complete basis set (CBS) limit with a polynomial convergence rate, 
which can be exploited to reduce the remaining basis set incompleteness (BSI) error via
 basis set extrapolation approaches~\cite{KMSCFCBS,CorrConv}.  
We employ two-point formulae for extrapolation, labeled as CBS($X$,$X+1$), where $X$ refers 
to the cardinal number of the  aug-cc-pV$X$Z basis set~\cite{AUGPVXZ} with $X$=T, Q, and 5.

 For the proper description of important medium- and long-range
interactions and of the cross-polarization of the monomers in the complex,
it is crucial to the employ diffuse, i.e., spatially spread basis functions. 
The use of such diffuse basis functions, however, greatly enhances technical challenges characteristic 
of interaction energy computations with atom centered Gaussian type basis functions. 
As long as the 
basis set expansion of the monomers is not saturated completely, the basis functions residing on the atoms of one
monomer can contribute to the description of the wavefunction components of the other monomer. 
Thus, the resulting basis set superposition error (BSSE) emerges from the 
unbalanced improvement of the basis set expansion of the monomers and the dimer and usually leads to 
artificially 
overestimated interaction strength.
The BSSE can be decreased 
significantly by counterpoise (CP) corrections~\cite{BSSE}, i.e., by using the entire 
dimer basis set also for the monomer calculations. Naturally, for small basis sets 
this approach might lead to a more saturated basis set expansion on the monomers and 
can potentially overcorrect the BSSE. In the case of 
aug-cc-pV$X$Z with $X$=T, Q, and 5 the CP correction decreases monotonically 
with increasing basis set size, thus a decreasing CP correction is an excellent indicator of 
basis set saturation, which we employ here.

To characterize the convergence of the LNO-CCSD(T) interaction energies in terms of the basis set 
completeness, the maximum of two BSI error indicators is considered with the best available 
LNO threshold set. One of them is the difference of the
CP corrected and uncorrected LNO-CCSD(T)/CBS(Q,5) interaction energies, and the other one is the difference of CP corrected 
 LNO-CCSD(T)/CBS(T,Q) and LNO-CCSD(T)/CBS(Q,5) interaction energies.
 The resulting 
BSI error bar values of Table~\ref{lccres} indicate that 
the above two four-membered groups exhibit much more
homogeneous basis set convergence behavior. 
For the GGG, GCGC, PHE, and CBH interaction energies, this BSI measure is 0.06-0.22 kcal mol$^{-1}$, while 
for the other four complexes a twice as large uncertainty of 0.19-0.36 kcal mol$^{-1}$ is found.
Compared to the similar or larger local error bars, 
we find this 
level of basis set convergence to be highly satisfactory. 

We again investigate more closely only the C3A, C3GC, C2C2PD, and C$_{60}$@[6]CPPA quartet. 
The convergence of LNO-CCSD(T) interaction energies with improving basis sets for C3A and C3GC is shown on 
panel a) of Fig.~\ref{fig3:convergence} of the main text. The large BSSE obtained with 
the aug-cc-pVTZ, and to some extent also with the aug-cc-pVQZ basis set is apparent for both complexes.
Such large BSSE also affects the extrapolation, the CBS(T,Q) results clearly overshoot the basis set limit due to the 
underestimation of the aug-cc-pVTZ result. The BSSE is significantly reduced by the CP correction.
All CP corrected results (solid symbols) closely agree already at the aug-cc-pVTZ level. 
Most importantly, the CBS(Q,5) entries of both the CP corrected and uncorrected series match 
each other within a few tenth of a kcal mol$^{-1}$, providing strong indication of basis set saturation. 
Upon inspection of the CP corrected and uncorrected interaction energies of Table~\ref{lccres}, 
this statement can be extended for the remaining six complexes as well. 

The left and right panels of Fig.~\ref{coronene_basis} collect CP uncorrected and corrected 
LNO-CCSD(T) interaction energies for the coronene dimer. The overbinding of the aug-cc-pVTZ and 
aug-cc-pVQZ results caused by the BSSE is again significant, close to 50 and 20\%, respectively. 
With the exception of the overshooting CBS(T,Q) extrapolation,
the aug-cc-pV$X$Z energies, with $X$=T, Q, and 5, as well as the CBS(Q,5) extrapolation form a highly convincing, converging 
series of results both with and without CP correction. The CP corrected and uncorrected CBS
results approach the region of convergence from the opposite directions, hence their average, i.e., 
the half CP corrected results appear to be the best estimate at the CBS(Q,5) level.
Concerning CBS(T,Q), the fully CP corrected results are found more reliable
due to the excessive BSSE obtained with aug-cc-pVTZ. 

\pn{Although we find the level of convergence regarding the basis set satisfactory, we invested additional efforts 
to perform LNO-CCSD(T)/aug-cc-pV6Z computations for the GGG complex. The CP corrected CBS(Q,5) and CBS(5,6) results 
at the very Tight LNO-CCSD(T) level agree up to 0.1 kcal mol$^{-1}$, which is within the local uncertainty.}

Finally, we assess the accuracy of the composite BSI correction approach employed for C3A, C3GC, and C$_{60}$@[6]CPPA.
 Due to the prohibitive computational costs, the most accurate interaction 
energies presented here for these three systems are obtained by adding 
a $\Delta_\text{BSI} = E^{\mathrm{LNO-CCSD(T)}}_\mathrm{Normal}[\mathrm{CBS(Q,5)}]- E^{\mathrm{LNO-CCSD(T)}}_\mathrm{Normal}$[aug-cc-pVTZ] 
BSI correction to the 
 $E^{\mathrm{LNO-CCSD(T)}}_\mathrm{Tight-very \; Tight}$[aug-cc-pVTZ]
interaction energies. 
This formula exploits the similarity of the local approximation
convergence curves obtained with different basis sets and it is numerically identical to 
$E^{\mathrm{LNO-CCSD(T)}}_\mathrm{Tight-very \; Tight}$[CBS(Q,5)]
if the local convergence patterns are exactly parallel.
To assess the quality of $\Delta_\text{BSI}$, we compared $\Delta_\text{BSI}$ to the analogous
$\Delta_\text{BSI}^{\text{very Tight}}= E^{\mathrm{LNO-CCSD(T)}}_\mathrm{very \; Tight}[\mathrm{CBS(Q,5)}]-E^{\mathrm{LNO-CCSD(T)}}_\mathrm{very \; Tight}$[aug-cc-pVTZ] wherever it is available. 
For the system most similar with the above three, that is, for C2C2PD, the 
$|\Delta_\text{BSI}^{\text{very Tight}}-\Delta_\text{BSI}|$ value is about 0.12 kcal mol$^{-1}$. To account for the potentially
size-extensive nature of this unparallelity error, the final $\Delta_\text{BSI}$ error estimates of Table~\ref{lccres}
were obtained by scaling the 0.12 kcal mol$^{-1}$ with the ratio of the interaction energies of the given complex and C2C2PD. 
The ``Total error bar'' values of Table~\ref{lccres} also include this third, $\Delta_\text{BSI}$ 
related uncertainty estimate for these three complexes.

\pn{
Even more details can be learned observing the convergence of the total HF and the LNO-CCSD(T) correlation energies separately for the 
complexes and monomers (see Sect.~\ref{absenergies}). The HF total energies are converged to six significant digits at the CBS(Q,5) level, 
which translates into a highly convincing convergence level of  0.01 kcal mol$^{-1}$ regarding the HF part of the interaction energies. 
In other words, that BSI error estimates collected in Table~\ref{lccres} have negligible HF and sizable correlation contribution. 
Furthermore, the CP corrected interaction energies are converged up to this 0.01 kcal mol$^{-1}$ level already with the smallest, 
aug-cc-pVTZ basis set. 
As expected, the CCSD(T) correlation energies tend significantly more slowly to the CBS limit with the cardinal number, but the agreement 
of the CBS(T,Q) and CBS(Q,5) values up to 4 significant digits is again highly satisfactory. This shows that the BSI error estimates being below 0.36 kcal mol$^{-1}$, just as 
the LNO error estimates, are consistent with the absolute energies and do not benefit from sizable error compensation. Additionally, 
 the computation of the interaction energies is warranted  according to the supermolecular approach [see Eq. (1) of the main text], because
total energies are converged to the necessary number of significant digits.
}

\pn{
\subsection{LNO-CCSD(T) energies plotted on Figs.~\ref{fig3:convergence}. and~ \ref{coronene_basis}} \label{absenergies}

In Tables~\ref{c3gctab}--\ref{c2tab}, we collect the absolute HF, the LNO-CCSD(T) correlation, and the corresponding 
interaction energies using all possible combinations of settings ({\it Normal} to {\it very Tight}, aug-cc-pVTZ to aug-cc-pV5Z, corresponding 
extrapolated energies, and various use of CP corrections) to document the numerical data 
plotted in Fig.~\ref{fig3:convergence} and~\ref{coronene_basis}. Additional analysis is provided in Sects.~\ref{localconvg} and ~\ref{CCbasis}.

\begin{table}[h!]
\scriptsize
\caption{HF energies and LNO-CCSD(T) correlation energies [in a.u.], and corresponding interaction energies [$\Delta$E in kcal mol$^{-1}$] obtained for the C3GC dimer with all employed 
basis sets and LNO threshold combinations, including CBS and LNO extrapolations as well as full (CP)  and half CP (half CP) corrected results.$^\text{a}$    
\label{c3gctab} }
\begin{tabular}{l|rrr|rr|ccc}
                 & C3GC         & circ.        & GC          & circ.  CP  & GC CP    & \;\;\;\;\; $\Delta$E   \;\;\;\;   & $\Delta$E  CP & $\Delta$E  half CP \\ \hline
aug-cc-pVTZ      &              &              &             &              &             &        &          &               \\ 
HF               & -2988.683486 & -2056.327131 & -932.381768 & -2056.328307 & -932.383046 & 15.95  & 17.49    & 16.72         \\
Normal           & -12.6945     & -8.8704      & -3.7261     & -8.8783      & -3.7319     & -45.57 & -35.46   & -40.52        \\
Tight            & -12.6941     & -8.8746      & -3.7284     & -8.8824      & -3.7341     & -41.15 & -31.19   & -36.17        \\
very Tight       & -12.6955     & -8.8771      & -3.7294     & -8.8848      & -3.7352     & -39.84 & -29.90   & -34.87        \\
Normal--Tight     & -12.6938     & -8.8768      & -3.7296     & -8.8844      & -3.7353     & -38.93 & -29.05   & -33.99        \\
Tight--very Tight & -12.6962     & -8.8784      & -3.7299     & -8.8860      & -3.7357     & -39.19 & -29.25   & -34.22        \\ \hline
aug-cc-pVQZ      &              &              &             &              &             &        &          &               \\ 
HF               & -2988.845437 & -2056.435660 & -932.437072 & -2056.435922 & -932.437370 & 17.13  & 17.48    & 17.30         \\
Normal           & -13.2457     & -9.2508      & -3.9073     & -9.2529      & -3.9092     & -37.84 & -34.95   & -36.40        \\
Tight            & -13.2452     & -9.2551      & -3.9096     & -9.2570      & -3.9115     & -33.42 & -30.67   & -32.05        \\
very Tight       & -13.2467     & -9.2576      & -3.9106     & -9.2595      & -3.9125     & -32.11 & -29.38   & -30.75        \\
Normal--Tight     & -13.2450     & -9.2572      & -3.9108     & -9.2591      & -3.9126     & -31.20 & -28.54   & -29.87        \\ 
Tight--very Tight & -13.2474     & -9.2588      & -3.9111     & -9.2607      & -3.9130     & -31.46 & -28.74   & -30.10        \\ \hline
aug-cc-pV5Z      &              &              &             &              &             &        &          &               \\ 
HF               & -2988.878943 & -2056.457991 & -932.448732 & -2056.458027 & -932.448766 & 17.43  & 17.48    & 17.45         \\ 
Normal           & -13.4437     & -9.3858      & -3.9722     & -9.3874      & -3.9728     & -36.35 & -34.95   & -35.65        \\
Tight            & -13.4432     & -9.3901      & -3.9745     & -9.3914      & -3.9751     & -31.92 & -30.67   & -31.30        \\
very Tight       & -13.4446     & -9.3926      & -3.9755     & -9.3939      & -3.9761     & -30.62 & -29.38   & -30.00        \\
Normal--Tight     & -13.4430     & -9.3922      & -3.9757     & -9.3935      & -3.9762     & -29.71 & -28.54   & -29.12        \\
Tight--very Tight & -13.4453     & -9.3938      & -3.9760     & -9.3951      & -3.9766     & -29.97 & -28.74   & -29.35        \\  \hline
CBS(T,Q)         &              &              &             &              &             &        &          &               \\
HF               & -2988.889784 & -2056.465378 & -932.452216 & -2056.465391 & -932.452245 & 17.45  & 17.48    & 17.46         \\
Normal           & -13.6479     & -9.5285      & -4.0395     & -9.5263      & -4.0387     & -32.74 & -34.57   & -33.66        \\
Tight            & -13.6475     & -9.5327      & -4.0418     & -9.5304      & -4.0409     & -28.31 & -30.30   & -29.31        \\
very Tight       & -13.6489     & -9.5352      & -4.0428     & -9.5329      & -4.0420     & -27.01 & -29.01   & -28.01        \\
Normal--Tight     & -13.6472     & -9.5349      & -4.0430     & -9.5325      & -4.0420     & -26.10 & -28.16   & -27.13        \\
Tight--very Tight & -13.6496     & -9.5365      & -4.0433     & -9.5341      & -4.0425     & -26.36 & -28.36   & -27.36        \\ \hline
CBS(Q,5)         &              &              &             &              &             &        &          &               \\
HF               & -2988.884505 & -2056.461698 & -932.450668 & -2056.461696 & -932.450658 & 17.48  & 17.48    & 17.48         \\
Normal           & -13.6514     & -9.5274      & -4.0403     & -9.5284      & -4.0395     & -35.05 & -34.94   & -35.00        \\
Tight            & -13.6509     & -9.5317      & -4.0426     & -9.5325      & -4.0417     & -30.63 & -30.67   & -30.65        \\
very Tight       & -13.6524     & -9.5342      & -4.0436     & -9.5349      & -4.0428     & -29.32 & -29.38   & -29.35        \\
Normal--Tight     & -13.6507     & -9.5338      & -4.0438     & -9.5345      & -4.0429     & -28.42 & -28.53   & -28.47        \\
Tight--very Tight & -13.6531     & -9.5354      & -4.0441     & -9.5361      & -4.0433     & -28.67 & -28.73   & -28.70       
\end{tabular}
\\ \textsuperscript{\emph{a}} Tight and very Tight results obtained with aug-cc-pVQZ and aug-cc-pV5Z, as well as any derivatives thereof
 employ the additive BSI correction according to the Methods and ~\ref{CCbasis} Sections. 
\end{table}

\begin{table}[h!]
\scriptsize
\caption{HF and LNO-CCSD(T) energies for systems of the C3A complex. See caption of~\ref{c3gctab} for more details.
\label{c3atab} }
\begin{tabular}{l|rrr|rr|ccc}
                 & C3A          & circ.        & adenine     & circ. CP  & adenine  CP & \;\;\;\;\; $\Delta$E   \;\;\;\;   & $\Delta$E CP & $\Delta$E half CP \\ \hline
aug-cc-pVTZ      &              &              &             &              &               &        &          &               \\
HF               & -2520.996126 & -2056.327134 & -464.682371 & -2056.327896 & -464.683038   & 8.40   & 9.29     & 8.84          \\
Normal           & -10.8264     & -8.8705      & -1.9001     & -8.8753      & -1.9033       & -26.65 & -20.77   & -23.71        \\
Tight            & -10.8272     & -8.8746      & -1.9008     & -8.8792      & -1.9038       & -24.12 & -18.40   & -21.26        \\
very Tight       & -10.8285     & -8.8770      & -1.9009     & -8.8818      & -1.9040       & -23.32 & -17.52   & -20.42        \\
Normal--Tight     & -10.8276     & -8.8767      & -1.9011     & -8.8812      & -1.9041       & -22.86 & -17.22   & -20.04        \\
Tight--very Tight & -10.8291     & -8.8783      & -1.9010     & -8.8831      & -1.9040       & -22.92 & -17.08   & -20.00        \\ \hline
aug-cc-pVQZ      &              &              &             &              &               &        &          &               \\
HF               & -2521.130595 & -2056.435693 & -464.709364 & -2056.435867 & -464.709527   & 9.08   & 9.29     & 9.18          \\
Normal           & -11.2905     & -9.2505      & -1.9900     & -9.2518      & -1.9912       & -22.27 & -20.53   & -21.40        \\
Tight            & -11.2912     & -9.2546      & -1.9907     & -9.2558      & -1.9917       & -19.74 & -18.16   & -18.95        \\
very Tight       & -11.2925     & -9.2571      & -1.9908     & -9.2584      & -1.9918       & -18.94 & -17.27   & -18.11        \\
Normal--Tight     & -11.2916     & -9.2567      & -1.9910     & -9.2577      & -1.9920       & -18.48 & -16.97   & -17.73        \\
Tight--very Tight & -11.2932     & -9.2583      & -1.9909     & -9.2597      & -1.9919       & -18.54 & -16.83   & -17.69        \\ \hline
aug-cc-pV5Z      &              &              &             &              &               &        &          &               \\
HF               & -2521.158239 & -2056.458030 & -464.714962 & -2056.458051 & -464.714981   & 9.26   & 9.28     & 9.27          \\
Normal           & -11.4564     & -9.3855      & -2.0221     & -9.3867      & -2.0225       & -21.34 & -20.26   & -20.80        \\
Tight            & -11.4571     & -9.3897      & -2.0227     & -9.3907      & -2.0231       & -18.81 & -17.89   & -18.35        \\
very Tight       & -11.4584     & -9.3921      & -2.0229     & -9.3933      & -2.0232       & -18.01 & -17.01   & -17.51        \\
Normal--Tight     & -11.4575     & -9.3917      & -2.0231     & -9.3927      & -2.0234       & -17.54 & -16.70   & -17.12        \\
Tight--very Tight & -11.4591     & -9.3933      & -2.0229     & -9.3946      & -2.0233       & -17.61 & -16.56   & -17.09        \\ \hline
CBS(T,Q)         &              &              &             &              &               &        &          &               \\
HF               & -2521.167416 & -2056.465420 & -464.716755 & -2056.465433 & -464.716780   & 9.26   & 9.29     & 9.27          \\
Normal           & -11.6291     & -9.5278      & -2.0556     & -9.5266      & -2.0553       & -19.38 & -20.35   & -19.87        \\
Tight            & -11.6299     & -9.5320      & -2.0563     & -9.5305      & -2.0559       & -16.86 & -17.98   & -17.42        \\
very Tight       & -11.6312     & -9.5344      & -2.0564     & -9.5331      & -2.0560       & -16.06 & -17.09   & -16.57        \\
Normal--Tight     & -11.6303     & -9.5340      & -2.0566     & -9.5325      & -2.0562       & -15.59 & -16.79   & -16.19        \\
Tight--very Tight & -11.6318     & -9.5356      & -2.0565     & -9.5344      & -2.0560       & -15.66 & -16.65   & -16.15        \\ \hline
CBS(Q,5)         &              &              &             &              &               &        &          &               \\
HF               & -2521.162828 & -2056.461737 & -464.715891 & -2056.461734 & -464.715887   & 9.29   & 9.28     & 9.28          \\
Normal           & -11.6304     & -9.5272      & -2.0557     & -9.5283      & -2.0555       & -20.52 & -19.97   & -20.25        \\
Tight            & -11.6312     & -9.5313      & -2.0564     & -9.5323      & -2.0561       & -17.99 & -17.60   & -17.80        \\
very Tight       & -11.6324     & -9.5338      & -2.0565     & -9.5349      & -2.0562       & -17.19 & -16.72   & -16.95        \\
Normal--Tight     & -11.6315     & -9.5334      & -2.0567     & -9.5342      & -2.0563       & -16.73 & -16.42   & -16.57        \\
Tight--very Tight & -11.6331     & -9.5350      & -2.0566     & -9.5362      & -2.0562       & -16.79 & -16.28   & -16.53       
\end{tabular}
\end{table}

\begin{table}[h!]
\scriptsize
\caption{HF and LNO-CCSD(T) energies for systems of the C2C2PD complex. See caption of \ref{c3gctab} for more details.$^\text{a}$
\label{c2tab} }
\begin{tabular}{l|rrr|ccc}
                 & C2C2PD       & \;\;\;\;\; coronene \;\;\;\;\;   & coronene CP & \;\;\;\;\; $\Delta$E   \;\;\;\;   & $\Delta$E CP & $\Delta$E half CP \\ \hline
aug-cc-pVTZ      &              &             &                &        &          &               \\
HF               & -1832.429002 & -916.226325 & -916.227229    & 14.84  & 15.97    & 15.41         \\
Normal           & -8.0606      & -3.9915     & -3.9968        & -33.84 & -26.01   & -29.92        \\
Tight            & -8.0585      & -3.9929     & -3.9982        & -30.76 & -22.93   & -26.84        \\
very Tight       & -8.0574      & -3.9932     & -3.9985        & -29.68 & -22.01   & -25.84        \\
Normal--Tight     & -8.0574      & -3.9936     & -3.9989        & -29.22 & -21.39   & -25.30        \\
Tight--very Tight & -8.0569      & -3.9934     & -3.9986        & -29.14 & -21.55   & -25.34        \\ \hline
aug-cc-pVQZ      &              &             &                &        &          &               \\
HF               & -1832.526572 & -916.275811 & -916.276010    & 15.72  & 15.97    & 15.84         \\
Normal           & -8.3972      & -4.1641     & -4.1655        & -27.58 & -25.62   & -26.60        \\
Tight            & -8.3898      & -4.1628     & -4.1641        & -24.57 & -22.66   & -23.62        \\
very Tight       & -8.3866      & -4.1621     & -4.1635        & -23.49 & -21.43   & -22.46        \\
Normal--Tight     & -8.3861      & -4.1622     & -4.1635        & -23.07 & -21.19   & -22.13        \\
Tight--very Tight & -8.3850      & -4.1617     & -4.1632        & -22.94 & -20.82   & -21.88        \\ \hline
aug-cc-pV5Z      &              &             &                &        &          &               \\
HF               & -1832.546868 & -916.286130 & -916.286155    & 15.93  & 15.97    & 15.95         \\
Normal           & -8.5176      & -4.2252     & -4.2258        & -26.24 & -25.44   & -25.84        \\
Tight            & -8.5008      & -4.2194     & -4.2199        & -22.98 & -22.33   & -22.66        \\
very Tight       & -8.4937      & -4.2166     & -4.2171        & -22.05 & -21.33   & -21.69        \\
Normal--Tight     & -8.4924      & -4.2165     & -4.2169        & -21.35 & -20.78   & -21.07        \\
Tight--very Tight & -8.4901      & -4.2152     & -4.2158        & -21.59 & -20.83   & -21.21        \\ \hline
CBS(T,Q)         &              &             &                &        &          &               \\
HF               & -1832.553290 & -916.289361 & -916.289368    & 15.96  & 15.97    & 15.96         \\
Normal           & -8.6429      & -4.2901     & -4.2886        & -23.42 & -25.34   & -24.38        \\
Tight            & -8.6316      & -4.2868     & -4.2852        & -20.46 & -22.47   & -21.46        \\
very Tight       & -8.6268      & -4.2852     & -4.2839        & -19.37 & -21.01   & -20.19        \\
Normal--Tight     & -8.6260      & -4.2852     & -4.2835        & -18.98 & -21.03   & -20.00        \\
Tight--very Tight & -8.6243      & -4.2845     & -4.2833        & -18.83 & -20.28   & -19.55        \\ \hline
CBS(Q,5)         &              &             &                &        &          &               \\
HF               & -1832.550237 & -916.287842 & -916.287839    & 15.97  & 15.96    & 15.97         \\
Normal           & -8.6439      & -4.2893     & -4.2891        & -25.02 & -25.25   & -25.13        \\
Tight            & -8.6172      & -4.2788     & -4.2784        & -21.50 & -21.98   & -21.74        \\
very Tight       & -8.6061      & -4.2738     & -4.2734        & -20.74 & -21.22   & -20.98        \\
Normal--Tight     & -8.6039      & -4.2735     & -4.2730        & -19.74 & -20.35   & -20.05        \\ 
Tight--very Tight & -8.6005      & -4.2713     & -4.2709        & -20.36 & -20.84   & -20.60       
\end{tabular}
\\ \textsuperscript{\emph{a}} Tight and very Tight results obtained with aug-cc-pVQZ and aug-cc-pV5Z are also directly evaluated 
without replying on the additive BSI correction according to the Methods and \ref{CCbasis} Sections.
\end{table}
}

\pn{
\clearpage
\subsection{Core and higher-order correlation on top of CCSD(T)} \label{coreTQ}

Core correlation effects are evaluated using the highly-optimized density-fitting (DF) MP2 implementation of the {\sc Mrcc} package~\cite{MRCC}
using large basis sets. In that way, the magnitude of the frozen core approximation can be determined independently from the local and BSI errors. 
The augmented weighted core-valence basis sets~\cite{wCVXZ12Row}, aug-cc-pwCV$X$Z with $X$=T and Q, were employed in combination with CP corrections. 
The core correlated DF-MP2 interaction energies of the C2C2PD complex, both with aug-cc-pwCVTZ and with aug-cc-pwCVQZ, as well as with CBS(T,Q),
are consistently stronger by 0.22~kcal~mol$^{-1}$ (4.6~cal~mol$^{-1}$ per C atom)
than those obtained using the frozen core approach and otherwise 
identical settings. 

The missing higher-order electron correlation on top of the CCSD(T) treatment was estimated using the 
CCSDT(Q) scheme, which includes 
infinite-order three-electron and the perturbative four-electron contributions~\cite{Pert}.
As the conventional ninth power-scaling CCSDT(Q) calculations are many-orders of magnitude more expensive than CCSD(T),
we relied on the analogous LNO approximations implemented also for CCSDT(Q)~\cite{LocalCC,LocalCC3} in the {\sc Mrcc} package~\cite{MRCC}.
Considering that large basis set CCSDT(Q) computations are only feasible for systems with only a few atoms, 
highly-converged LNO-CCSDT(Q) computations are still well beyond the current capabilities even for the smallest GGG complex.
With relying on looser LNO truncations and the moderate basis set of
6-31G**(0.25,0.15)~\cite{L7set}, we were able to perform by far the largest LNO-CCSDT(Q) calculation ever presented for the GGG complex.
The cumulative local and BSI error of the LNO-CCSDT(Q) interaction energies are estimated to be about 38\% 
at the corresponding LNO-CCSD(T)/6-31G**(0.25,0.15) level.
Up to this uncertainty, the CCSDT contribution on top of CCSD(T) is found to be -0.013~kcal~mol$^{-1}$, while 
the (Q) correction on top of CCSDT is about -0.021~kcal~mol$^{-1}$. Clearly, both corrections are negligibly small compared to 
the deviation of CCSD(T) and FN-DMC. As the even higher-order CC terms are expected to 
be even smaller, it is unlikely that higher-order electron correlation effects missing from CCSD(T) could completely 
explain the disagreement of CCSD(T) and FN-DMC.

The weakly-correlated character of the studied system is also verified via the T1~\cite{t1measure} diagnostics. The T1 measures obtained 
for the most complicated C3GC and C$_{60}$@[6]CPPA complexes are found to be at most 0.016 and 0.014, respectively. 
Considering that the T1 measure grows with the number of basis functions and that smaller than 0.02 T1 values are considered 
weakly-correlated already for very small systems~\cite{t1measure}, there appears to be no indication of even moderate static correlation. 
Moreover, neither the HF  nor the CCSD iterations indicated any problems emerging usually for strongly correlated systems. 
The size of the singles and doubles amplitudes were also monitored in all domain CCSD computations indicating the validity of the single-reference approach,
while it is convincing that  the LNO approximations were found to operate excellently also for moderately statically correlated species~\cite{polypyrrolBM}.
The magnitude of the (T) correction compared to the full CCSD(T) interaction energy is also an informative measure of the static or dynamic nature of the correlation.
 This ratio is consistently around 18-20\% for all 8 complexes, which is well within the range  observed for smaller and simpler systems, e.g., in the well-known S66 test set (cca. 13--24\%)~\cite{S66}.
}

\section{Details of Quantum Monte Carlo calculations} \label{SMDMC}
The FN-DMC calculations mostly used 10 nodes with 28 cores each, and 14,000 walkers distributed across the cores (\textit{i.e.} 50 walkers per core). We used 20 nodes for the C$_{60}$@[6]CPPA complex and 28,000 walkers to reduce the stochastic error in a shorter time. Here we give further details on (i) the optimization of the Jastrow factor for the reported complexes, (ii) time-step and node-structure tests for the coronene dimer and (iii) results of additional FN-DMC simulations of C$_{60}$@[6]CPPA. In addition, we report the total energies for C3A and C3GC complexes here in Table~\ref{tab:dmcabsenergy}.

\begin{table}[ht]
\centering
\caption{The total energy in Ha for each complex and its monomers are given here with stochastic errors from FN-DMC calculations alongside description of the DFT orbitals and plane-wave cut-offs (in Ry) and FN-DMC time step (in a.u.) and algorithm (\textit{i.e.} standard locality approximation (LA) or determinant localization approximation (DLA)). Resulting interaction energy (IE) is reported lastly.}
\label{tab:dmcabsenergy}
\scriptsize
\begin{tabular}{ccccc}
\hline \hline
FN-DMC setup & C3GC & circ. & GC & IE (kcal mol$^{-1}$)\\ \hline
LDA orb/500Ry, 0.03 time-step/LA   & $-485.928809\pm0.000934$ & $-317.320337\pm0.000485$ & $-168.569987\pm0.000209$ & $-24.2 \pm 0.7$  \\
PBE0 orb/400Ry, 0.03 time-step/LA  & $-485.939114\pm0.001047$ & $-317.326450\pm0.000731$ & $-168.575218\pm0.000148$ & $-23.5 \pm 0.8$\\
PBE0 orb/400Ry, 0.01 time-step/LA  & $-485.933462\pm0.000881$ & $-317.325540\pm0.000579$ & $-168.570123\pm0.000603$ & $-23.8 \pm 0.8$ \\
\hline 
FN-DMC setup & C3A & circ. & adenine & IE (kcal mol$^{-1}$)\\ \hline
LDA orb/500Ry, 0.03 time-step/LA   & $-398.351177\pm0.000648$ & $-317.320307\pm0.000492$ & $-81.006922\pm0.000162$  & $-15.0 \pm 0.5$ \\ 
LDA orb/500Ry, 0.01 time-step/LA   & $-398.351280\pm0.000966$ & $-317.321748\pm0.000886$ & $-81.006670\pm0.000196$  & $-14.3 \pm 0.8$ \\
LDA orb/500Ry, 0.01 time-step/DLA  & $-398.438355\pm0.000546$ & $-317.393190\pm0.000585$ & $-81.022893\pm0.000228$  & $-14.0 \pm 0.5$ \\
\hline \hline 
\end{tabular}%
\end{table}

\subsection{Variational Monte Carlo Optimization of the Jastrow Factor}

Variational Monte Carlo (VMC) obeys the variational principle, allowing the initial Slater-Jastrow wavefunction to be optimized iteratively towards a lower energy. Importantly, the zero-variance principle 
ensures that variance of the energy tends to zero as the exact energy of the system is approached. This is used in the varmin and varmin--linjas optimization algorithms in CASINO~\cite{Needs_2009} to optimize the variable parameters of the Jastrow factor. The Jastrow factor is composed of explicit distance-dependent polynomial functions for inter-particle interactions, such as electron-electron (\textit{u}), electron-nucleus ($\chi$), and electron-electron-nucleus (\textit{f}), and is also system-dependent. For all complexes, we performed a term-by-term optimization using 24 parameters for \textit{u}, 12-14 parameters per element for $\chi$, and 8 parameters per element for \textit{f}. The resulting VMC energy and variance for the complexes is given in Table~\ref{tab:vmcopt}.
\begin{table}[ht]
\centering
\caption{The variance ($\sigma^2$) and VMC energy ($\textrm{E}_{\textrm{VMC}}$) in atomic units for each complex, as a result of optimizing the trial wavefunction. The uncertainty is indicated in parentheses.}
\label{tab:vmcopt}
\small
\begin{tabular}{ccc}
\hline \hline
Complex     & $\sigma^2$ & $\textrm{E}_{\textrm{VMC}}$ \\ \hline
CBH         & 4.07(5)       & -249.26(1)   \\
C2C2PD      & 4.76(6)       & -285.769(8)  \\
GGG         & 5.18(3)       & -290.195(5)  \\
GCGC        & 5.71(3)       & -336.296(1)  \\
PHE         & 6.03(4)       & -367.239(8)  \\
C3A         & 6.38(4)       & -397.085(6)  \\
C3GC        & 7.82(5)       & -484.474(7)  \\
C$_{60}$@[6]CPPA & 10.54(5)      & -624.140(6) \\ \hline \hline 
\end{tabular}%
\end{table}

\subsection{Time-Step and Node-Structure Dependence of the Coronene Dimer}

\begin{figure}[ht]
\includegraphics[width=0.45\textwidth]{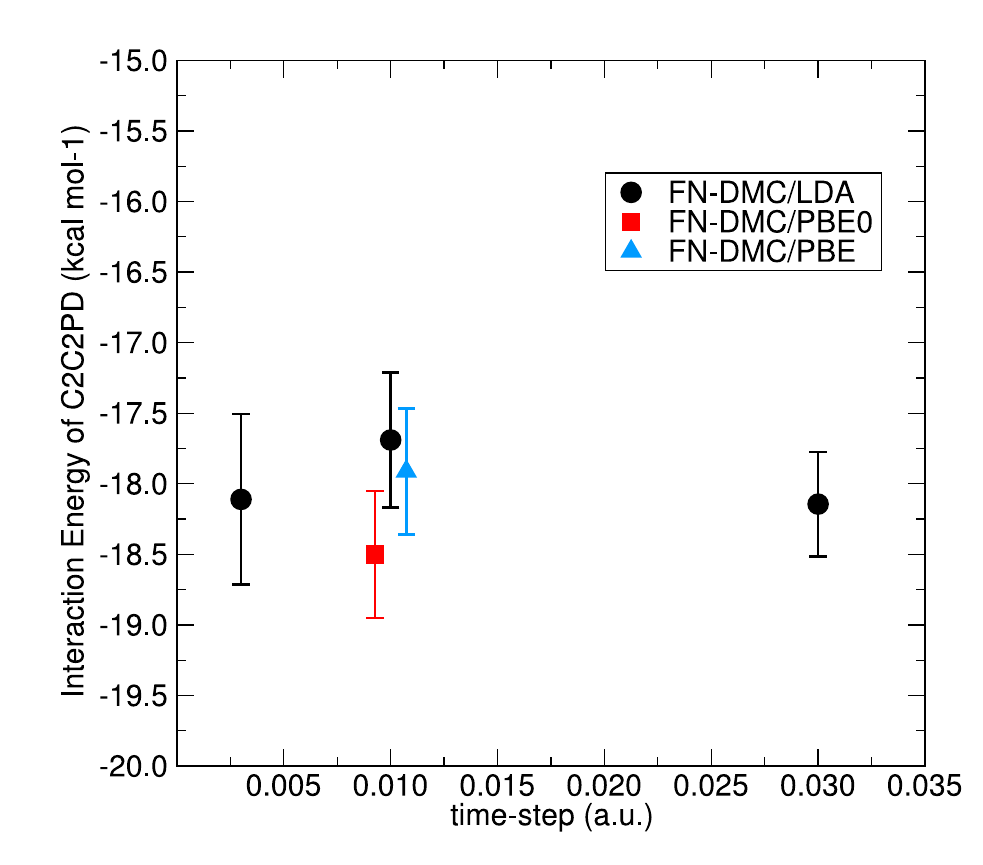}
\caption{\label{fig:convc2c2pd} FN-DMC interaction energy of C2C2PD (coronene dimer) with 0.003, 0.01 and 0.03 a.u. time-steps. Different nodal-structures from LDA (black circle), PBE (blue triangle), and PBE0 (red square) initial orbitals are reported using 0.01 a.u. time-step; these are slightly offset along the x-axis for clarity. }
\end{figure}

It can be seen from Fig.~\ref{fig:convc2c2pd} that the FN-DMC interaction energy of C2C2PD is converged within the stochastic error bar (corresponding to 1 standard deviation) with respect to the time-step in FN-DMC (from 0.003 to 0.03 a.u.). In addition, we computed PBE0 and PBE initial 
determinants (orbitals) from PWSCF, in order to assess the FN-DMC dependence of the interaction energy on the nodal-structure. Fig.~\ref{fig:convc2c2pd} shows that the FN-DMC interaction energy is the same within the stochastic error bars of $\sim$0.5 kcal mol$^{-1}$ across the three nodal-structures. 

\subsection{The GGG Trimer and Coronene Dimer with the Determinant Localization Approximation}

Using non-local pseudopotentials in FN-DMC requires an approximation for the evaluation of the local energy -- not to be confused with the type of local approximations, such as LNO, made in local CCSD(T) methods. The recent determinant localization approximation (DLA) introduced by Zen \textit{et al.}~\cite{Zen2019} has some advantages over the pre-existing standard algorithms: the locality approximation~\cite{Mitas1991} (LA) and T-move scheme~\cite{Casula2006}). The DLA FN-DMC energies are less sensitive to the Jastrow factor that is used in combination with pseudopotentials at larger time-steps. This enables better overall convergence with the time-step in FN-DMC and the DLA method is also more numerically stable than LA. We tested the use of the DLA method for the GGG trimer and the coronene dimer and present the results in Table~\ref{tab:dla}.
\begin{table}[ht]
\centering
\caption{Comparison of the standard LA to the DLA method in the GGG and C2C2PD complexes.}
\label{tab:dla}
\small
\begin{tabular}{cccc}
\hline \hline
Complex & Approximation   & Time-step & IE (kcal mol$^{-1}$) \\ \hline
GGG & standard LA & 0.03 & $1.5\pm0.3$ \\
GGG & DLA & 0.03 & $1.4\pm0.2$ \\
C2C2PD & standard LA & 0.03 & $-18.1\pm0.4$ \\
C2C2PD & DLA & 0.01 & $-17.4\pm0.5$ \\ \hline \hline 
\end{tabular}%
\end{table}
The interaction energies of the GGG and C2C2PD complexes remain in agreement, within the one-standard deviation stochastic errors, between the DLA and the standard LA algorithms. The results support that the FN-DMC results are converged with respect to the time-steps and employed Jastrow factors. 

\subsection{FN-DMC with T-move on the C$_{60}$@[6]CPPA Complex}

\begin{figure}[ht]
\includegraphics[width=0.5\textwidth]{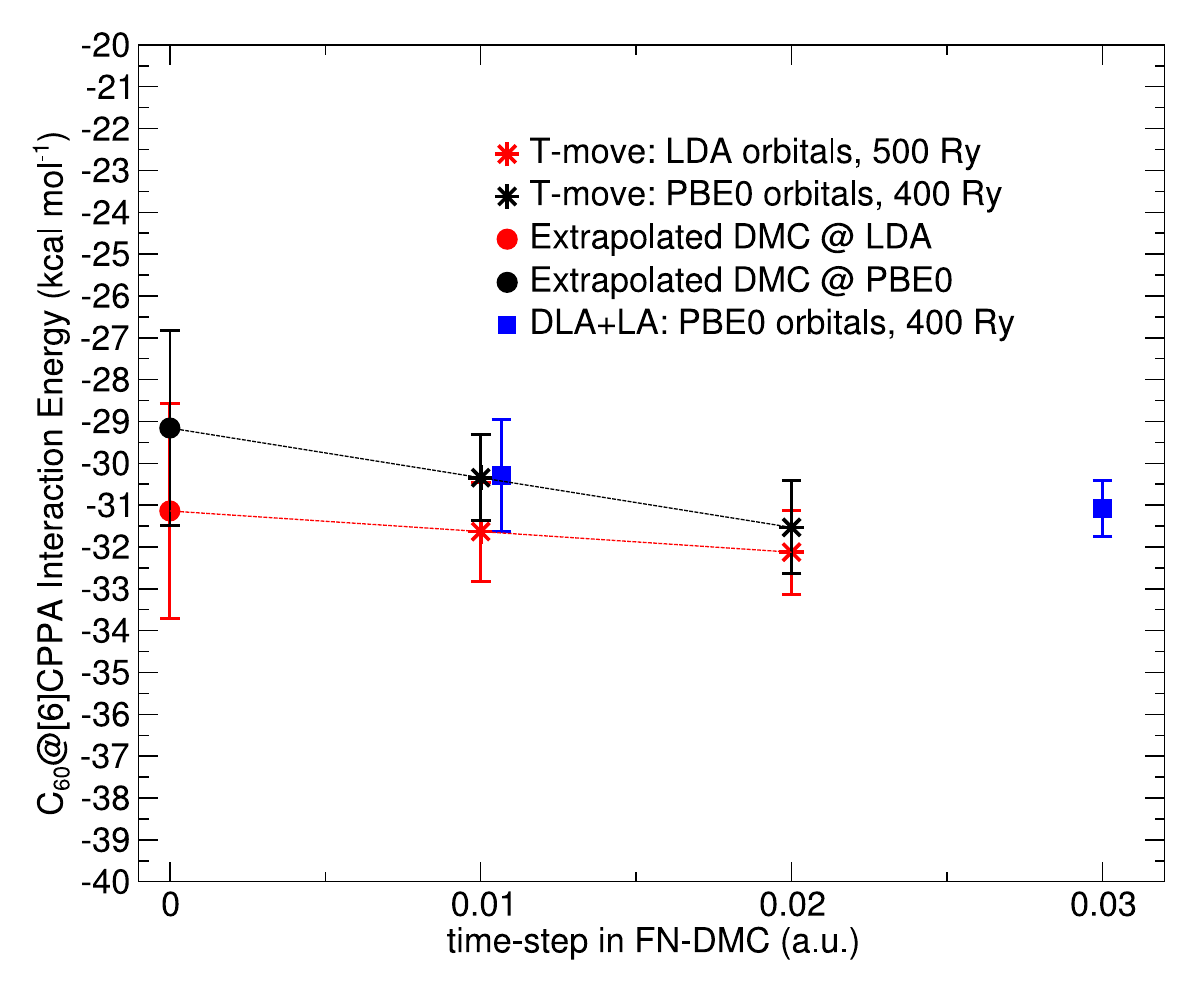}
\caption{\label{fig:convc606cppa} FN-DMC interaction energy of C$_{60}$@[6]CPPA complex using two algorithms. T-move interaction energies at 0.01 and 0.02 a.u. time-steps are shown for LDA (black stars) and PBE0 orbitals (red stars). The linear extrapolation to zero time-step for each set is indicated by the dashed lines, with the result in circles. The error on the zero time-step FN-DMC interaction energies are propagated according to the extrapolation. For comparison, the DLA method is shown (with the locality approximation) in blue squares. The DLA FN-DMC interaction energy at 0.01 a.u. is slightly offset along the x-axis for clarity. }
\end{figure}

The C$_{60}$@[6]CPPA complex proved to be more challenging to compute with FN-DMC, due to numerical instabilities when using the locality approximation. This was alleviated by the use of the DLA method, and separately using the T-move approximation in place of the locality approximation. The T-move scheme reinstates variational form of the energy, but the energies with this approximation are more time-step dependent, as can be seen in Fig.~\ref{fig:convc606cppa}. Linear extrapolations to 'zero' time-step limit yield $-31.14 \pm 2.57$ kcal mol$^{-1}$ using LDA orbitals and $-29.16 \pm 2.33$ kcal mol$^{-1}$ using PBE0 orbitals. Moreover, we show the DLA obtained FN-DMC interaction energy at 0.03 and 0.01 a.u. time-steps. In this way, the independence of the interaction energy on the nodal structure and the FN-DMC algorithm is established. 

\pn{
\section{Computational requirements of LNO-CCSD(T) and FN-DMC} \label{secmemtime}

CPU core time and minimal memory requirements are collected in Table \ref{memtime} for representative examples: the CBH and C3GC complexes. 
The very Tight LNO-CCSD(T)/aug-cc-pVTZ computation for C3GC was found to be the upper limit for the CPU time requirement among all LNO-CCSD(T) computations. 
Compared to that it is interesting to note the case of CBH, which contains even more atoms and almost as much AOs as C3GC. 
However, due to the relatively low complexity of the wavefunction of CBH, its CPU time demand is found to be up to 100 times smaller 
than that of C3GC when using the same settings. Unfortunately, the computations were scattered on multiple clusters and CPU types 
preventing the straightforward comparison of runtimes with various settings. For that reason CPU core times and the corresponding CPU types are reported.
With that in mind, we find similar trends as in our 
previous report~\cite{LocalCC4}. For instance, the memory requirement of LNO-CCSD(T) is exceptionally small compared to 
alternative CCSD(T) implementations, which was essential for the C3A, C3GC, and C$_{60}$@[6]CPPA computations. Moreover, about 
3-5 times more operations were performed when using one step tighter LNO settings, just as in our previous computations~\cite{LocalCC4},
 which trend is highly useful to estimate the feasibility of analogous computations. It is also important to note that the 
CPU and memory requirement grow much more slowly with the basis set size than with conventional CCSD(T), where the 
operation count and data storage increase by about a factor of 10 with one step in the cardinal number hierarchy (e.g., from aug-cc-pVTZ to aug-cc-pVQZ). 

Compared to LNO-CCSD(T), the FN-DMC runtimes depend less on the chemical 
composition and can be estimated more accurately based on the number of computed particles. The notably small memory requirement and the ease of efficient 
parallelization are also apparent benefits of the FN-DMC method. 
Moreover, the computational cost of FN-DMC does not change as steeply with various input nodal structures and time-steps allowing for the estimation 
of these effects using manageable additional computational time.
}

\begin{table}[h!]
\scriptsize
\label{memtime}
\caption{CPU core time (i.e., [number of nodes]*[core per node]*[wall time in years]) and minimum memory [in GB] requirement of the LNO-CCSD(T) and FN-DMC calculations for the CBH and C3GC complexes with all settings.}
\begin{tabular}{ll|cc|cc}
Complex: &  &   \multicolumn{2}{c}{CBH}  &   \multicolumn{2}{c}{C3GC} \\  \hline
No. of atoms:  & &  \multicolumn{2}{c}{112} & \multicolumn{2}{c}{101} \\
   & &  \;\;\; memory [GB]  \;\;\; & \;\;\;  time [core year] \;\;\;  & \;\;\; memory [GB]  \;\;\; & \;\;\;  time [core year] \;\;\; \\ \hline
LNO-CCSD(T)&AOs in aug-cc-pVTZ:      & \multicolumn{2}{c}{3404} & \multicolumn{2}{c}{4002} \\   
&Normal           & 3$^\text{\#}$       & 0.02$^\text{a}$           & 70$^\text{\#}$  & 0.4$^\text{a}$   \\
&Tight            & 7$^\text{\#}$       & 0.08$^\text{a}$           & 73$^\text{\#}$   &  3.6$^\text{a,b,e,*}$    \\
&very Tight       & 12$^\text{\#}$      & 0.2$^\text{a}$            & 200  & 20$^\text{c,g,*}$    \\ 
&AOs in aug-cc-pVQZ:   & \multicolumn{2}{c}{6376} & \multicolumn{2}{c}{7128} \\
&Normal           & 11$^\text{\#}$      & 0.04$^\text{a}$            & 110$^\text{\#}$  & 1.6$^\text{a,*}$      \\
&Tight            & 24$^\text{\#}$      & 0.13$^\text{a}$            & -  & -       \\
&very Tight       & 29$^\text{\#}$      & 0.4$^\text{d}$            & -  & -       \\ 
&AOs in aug-cc-pV5Z:     & \multicolumn{2}{c}{10652} & \multicolumn{2}{c}{11511} \\
&Normal           & 32$^\text{\#}$    & 0.1$^\text{e}$            & 63  & 2.7$^\text{f,*}$       \\
&Tight            & 50$^\text{\#}$    & 0.3$^\text{h}$            & -  & -       \\
&very Tight       & 86$^\text{\#}$       & 1.2$^\text{f,*}$        & -  & -       \\ \hline
FN-DMC & LA/0.03 a.u.     &     7$^{\dagger}$            &  2.5$^\text{c,e}$    &     15$^{\dagger}$     &   3.3$^\text{c,e}$ \\ 

\end{tabular}
\\ \textsuperscript{\emph{a}} Intel Xeon E5-2670 v3 2.3 GHz 
 \textsuperscript{\emph{b}}  Intel Xeon E5-1650 v2 3.5 GHz  
 \textsuperscript{\emph{c}} Intel Xeon Gold 6132 2.3 GHz    
 \textsuperscript{\emph{d}} Intel Xeon E5-2680 v2 2.8 GHz   
 \textsuperscript{\emph{e}} Intel Xeon E5-2680 v4 2.4 GHz   
 \textsuperscript{\emph{f}} Intel Xeon E5-2680 v3 2.5 GHz   
 \textsuperscript{\emph{g}} Intel Xeon Platinum 8180M 2.3 GHz   
 \textsuperscript{\emph{h}} Intel Xeon Gold 6138 1.9 GHz 
 \textsuperscript{\emph{*}} Estimated CPU time due to large number of restarts.
 \textsuperscript{\emph{\#}} Fully integral-direct integral transformation, minimal memory algorithm would require about up to 3--4 times less memory.
  \textsuperscript{\emph{$\dagger$}} Maximum shared memory used, mainly determined by size of wavefunction file.
\end{table}

\section{Details of DFT calculations}

The PBE0+MBD calculations were performed using FHI-aims v.190225 with all-electron numerical basis sets, with ``tight" defaults and tier 2 basis functions for all elements. The total energy threshold for self-consistent convergence was set to $10^{-7}$ eV. Spin and relativistic effects have not been included.
London dispersion energies from the D4 model are computed  with the {\sc dftd4} standalone program using the electronegativity equilibration charges (EEQ) and include a coupled-dipole based many-body dispersion correction (D4(EEQ)-MBD)~\cite{dftd4}.
The same geometries have been used as for the benchmark calculations for all structures. 

\section{Geometry of the L7 and the C$_{60}$@[6]CPPA complexes }

The structures and fragment definitions in Ref.~\citenum{L7set} were used for the L7 calculations. 
For C$_{60}$@[6]CPPA, a  C$_{70}$@[6]CPPA geometry from Ref.~\citenum{Hermann2017} was modified, by replacing C$_{70}$ with C$_{60}$ and the complex was symmetrized to D$_\text{3d}$ point group. The high-symmetry structure allows more efficient calculations with LNO-CCSD(T) with a speedup proportional to the rank of the point group~\cite{LocalCC,LocalCC3}. The stability of this complex was assessed by relaxing the geometry whilst retaining the symmetry group, at the DFT level (B97-3c exchange-correlation functional). The interaction strength increases by less than 0.1 kcal mol$^{-1}$ with respect to the unrelaxed structure. Relaxing the C$_{60}$ and [6]CPPA fragments reduces the interaction strength by 0.9 kcal mol$^{-1}$. 

The C$_{60}$@[6]CPPA Cartesian coordinates used in LNO-CCSD(T) and FN-DMC calculations is given here. 
\begin{center}
C       -0.72650728     -1.22225849     -3.24715547    \\
C        0.72650728     -1.22225849     -3.24715547 \\
C       -1.42176054     -0.01804451     -3.24715547 \\
C        1.42176054     -0.01804451     -3.24715547 \\
C        2.59727407      0.14217202     -2.40825772 \\
C        1.17551245     -2.32039134     -2.40825772 \\
C        2.30045705     -2.16706760     -1.60544793 \\
C        3.02696412     -0.90872044     -1.60544793 \\
C       -3.02696412     -0.90872044     -1.60544793 \\
C       -2.30045705     -2.16706760     -1.60544793 \\
C       -2.59727407      0.14217202     -2.40825772 \\
C       -1.17551245     -2.32039134     -2.40825772 \\
C        0.00000000     -2.99907290     -1.88979043 \\
C       -2.30045914     -2.68553418     -0.24808191 \\
C       -1.17551125     -3.33502315      0.24808191 \\
C        0.00000000     -3.49523729     -0.59081634 \\
C       -3.02696429      1.74761865     -0.59081634 \\
C       -3.47597040      0.64948897      0.24808191 \\
C       -2.59727332      1.49953645     -1.88979043 \\
C       -3.47597040     -0.64948897     -0.24808191 \\
C       -3.02696429     -1.74761865      0.59081634 \\
C       -3.02696412      0.90872044      1.60544793 \\
C       -2.59727407     -0.14217202      2.40825772 \\
C       -2.59727332     -1.49953645      1.88979043 \\
C       -0.72650707      3.07578804     -1.60544793 \\
C       -1.17551125      3.33502315     -0.24808191 \\
C       -1.42176162      2.17821932     -2.40825772 \\
C       -2.30045914      2.68553418      0.24808191 \\
C       -2.30045705      2.16706760      1.60544793 \\
C       -0.00000000      3.49523729      0.59081634 \\
C       -0.00000000      2.99907290      1.88979043 \\
C       -1.17551245      2.32039134      2.40825772 \\
C        0.69525326      1.24030300     -3.24715547 \\
C        1.42176162      2.17821932     -2.40825772 \\
C       -0.69525326      1.24030300     -3.24715547 \\
C        0.72650707      3.07578804     -1.60544793 \\
C        1.17551125      3.33502315     -0.24808191 \\
C        2.59727332      1.49953645     -1.88979043 \\
C        3.02696429      1.74761865     -0.59081634 \\
C        2.30045914      2.68553418      0.24808191 \\
C        0.72650728      1.22225849      3.24715547 \\
C        1.17551245      2.32039134      2.40825772 \\
C        2.30045705      2.16706760      1.60544793 \\
C        1.42176054      0.01804451      3.24715547 \\
C       -0.69525326     -1.24030300      3.24715547 \\
C       -1.42176054      0.01804451      3.24715547 \\
C       -0.72650728      1.22225849      3.24715547 \\
C        0.69525326     -1.24030300      3.24715547 \\
C        0.72650707     -3.07578804      1.60544793 \\
C       -0.72650707     -3.07578804      1.60544793 \\
C       -1.42176162     -2.17821932      2.40825772 \\
C        1.42176162     -2.17821932      2.40825772 \\
C        3.02696429     -1.74761865      0.59081634 \\
C        2.30045914     -2.68553418     -0.24808191 \\
C        1.17551125     -3.33502315      0.24808191 \\
C        2.59727332     -1.49953645      1.88979043 \\
C        3.02696412      0.90872044      1.60544793 \\
C        3.47597040      0.64948897      0.24808191 \\
C        3.47597040     -0.64948897     -0.24808191 \\
C        2.59727407     -0.14217202      2.40825772 \\
C       -4.43498968      4.84818396     -0.00326334 \\
C       -5.43089278      3.84285784     -0.00366147 \\
C       -3.85193459      5.28074380     -1.21899357 \\
C       -3.85171966      5.27950986      1.21280412 \\
C       -2.64632983      5.97544200     -1.21280412 \\
C       -2.64729099      5.97624511      1.21899357 \\
C       -1.98115563      6.26490571      0.00326334 \\
C       -6.04345890      2.78186219     -0.00366147 \\
H       -4.31742848      4.99255327     -2.16174586 \\
H       -4.31701338      4.99032575      2.15534928 \\
H       -2.16324219      6.23380613     -2.15534928 \\
H       -2.16496372      6.23527938      2.16174586 \\
C        1.98115563      6.26490571      0.00326334 \\
C        0.61256612      6.62472003      0.00366147 \\
C        2.64632983      5.97544200     -1.21280412 \\
C        2.64729099      5.97624511      1.21899357 \\
C        3.85193459      5.28074380     -1.21899357 \\
C        3.85171966      5.27950986      1.21280412 \\
C        4.43498968      4.84818396     -0.00326334 \\
C       -0.61256612      6.62472003      0.00366147 \\
H        2.16324219      6.23380613     -2.15534928 \\
H        2.16496372      6.23527938      2.16174586 \\
H        4.31742848      4.99255327     -2.16174586 \\
H        4.31701338      4.99032575      2.15534928 \\
C        6.41614531      1.41672175     -0.00326334 \\
C        6.04345890      2.78186219     -0.00366147 \\
C        6.49922558      0.69550131     -1.21899357 \\
C        6.49804949      0.69593214      1.21280412 \\
C        6.49804949     -0.69593214     -1.21280412 \\
C        6.49922558     -0.69550131      1.21899357 \\
C        6.41614531     -1.41672175      0.00326334 \\
C        5.43089278      3.84285784     -0.00366147 \\
H        6.48239220      1.24272611     -2.16174586 \\
H        6.48025557      1.24348038      2.15534928 \\
H        6.48025557     -1.24348038     -2.15534928 \\
H        6.48239220     -1.24272611      2.16174586 \\
C        4.43498968     -4.84818396      0.00326334 \\
C        5.43089278     -3.84285784      0.00366147 \\
C        3.85171966     -5.27950986     -1.21280412 \\
C        3.85193459     -5.28074380      1.21899357 \\
C        2.64729099     -5.97624511     -1.21899357 \\
C        2.64632983     -5.97544200      1.21280412 \\
C        1.98115563     -6.26490571     -0.00326334 \\
C        6.04345890     -2.78186219      0.00366147 \\
H        4.31701338     -4.99032575     -2.15534928 \\
H        4.31742848     -4.99255327      2.16174586 \\
H        2.16496372     -6.23527938     -2.16174586 \\
H        2.16324219     -6.23380613      2.15534928 \\
C       -1.98115563     -6.26490571     -0.00326334 \\
C       -0.61256612     -6.62472003     -0.00366147 \\
C       -2.64729099     -5.97624511     -1.21899357 \\
C       -2.64632983     -5.97544200      1.21280412 \\
C       -3.85171966     -5.27950986     -1.21280412 \\
C       -3.85193459     -5.28074380      1.21899357 \\
C       -4.43498968     -4.84818396      0.00326334 \\
C        0.61256612     -6.62472003     -0.00366147 \\
H       -2.16496372     -6.23527938     -2.16174586 \\
H       -2.16324219     -6.23380613      2.15534928 \\
H       -4.31701338     -4.99032575     -2.15534928 \\
H       -4.31742848     -4.99255327      2.16174586 \\
C       -6.41614531     -1.41672175      0.00326334 \\
C       -6.04345890     -2.78186219      0.00366147 \\
C       -6.49804949     -0.69593214     -1.21280412 \\
C       -6.49922558     -0.69550131      1.21899357 \\
C       -6.49922558      0.69550131     -1.21899357 \\
C       -6.49804949      0.69593214      1.21280412 \\
C       -6.41614531      1.41672175     -0.00326334 \\
C       -5.43089278     -3.84285784      0.00366147 \\
H       -6.48025557     -1.24348038     -2.15534928 \\
H       -6.48239220     -1.24272611      2.16174586 \\
H       -6.48239220      1.24272611     -2.16174586 \\
H       -6.48025557      1.24348038      2.15534928 \\
\end{center}



%

\end{document}